# Exomoon habitability constrained by illumination and tidal heating

René Heller[I] , Rory Barnes[II,III]

[I] Leibniz-Institute for Astrophysics Potsdam (AIP), An der Sternwarte 16, 14482 Potsdam, Germany, rheller@aip.de
[II] Astronomy Department, University of Washington, Box 951580, Seattle, WA 98195, rory@astro.washington.edu
[III] NASA Astrobiology Institute – Virtual Planetary Laboratory Lead Team, USA

## Abstract

The detection of moons orbiting extrasolar planets ("exomoons") has now become feasible. Once they are discovered in the circumstellar habitable zone, questions about their habitability will emerge. Exomoons are likely to be tidally locked to their planet and hence experience days much shorter than their orbital period around the star and have seasons, all of which works in favor of habitability. These satellites can receive more illumination per area than their host planets, as the planet reflects stellar light and emits thermal photons. On the contrary, eclipses can significantly alter local climates on exomoons by reducing stellar illumination. In addition to radiative heating, tidal heating can be very large on exomoons, possibly even large enough for sterilization. We identify combinations of physical and orbital parameters for which radiative and tidal heating are strong enough to trigger a runaway greenhouse. By analogy with the circumstellar habitable zone, these constraints define a circumplanetary "habitable edge". We apply our model to hypothetical moons around the recently discovered exoplanet Kepler-22b and the giant planet candidate KOI211.01 and describe, for the first time, the orbits of habitable exomoons. If either planet hosted a satellite at a distance greater than 10 planetary radii, then this could indicate the presence of a habitable moon.

## 1. Introduction

The question whether life has evolved outside Earth has prompted scientists to consider habitability of the terrestrial planets in the Solar System, their moons, and planets outside the Solar System, that is, extrasolar planets. Since the discovery of the first exoplanet almost two decades ago (Mayor & Queloz 1995), roughly 800 more have been found, and research on exoplanet habitability has culminated in the targeted space mission *Kepler*, specifically designed to detect Earth-sized planets in the circumstellar irradiation habitable zones (IHZs, Huang 1959; Kasting et al. 1993; Selsis et al. 2007; Barnes et al. 2009)[1] around Sun-like stars. No such Earth analog has been confirmed so far. Among the 2312 exoplanet candidates detected with *Kepler* (Batalha et al. 2012), more than 50 are indeed in the IHZ (Borucki et al. 2011; Kaltenegger & Sasselov 2011; Batalha et al. 2012), yet most of them are significantly larger and likely more massive than Earth. Habitability of the moons around these planets has received little attention. We argue here that it will be possible to constrain their habitability on the data available at the time they will be discovered.

Various astrophysical effects distinguish investigations of exomoon habitability from studies on exoplanet habitability. On a moon, there will be eclipses of the star by the planet (Dole 1964); the planet's stellar reflected light, as well the planet's thermal emission, might affect the moon's climate; and tidal heating can provide an additional energy source, which must be considered for evaluations of exomoon habitability (Reynolds et al. 1987; Scharf 2006; Debes & Sigurdsson 2007; Cassidy et al. 2009; Henning et al. 2009). Moreover, tidal processes between the moon and its parent planet will determine the orbit and spin evolution of the moon. Earth-sized planets in the IHZ around low-mass stars tend to become tidally locked, that is, one hemisphere permanently faces the star (Dole 1964; Goldreich 1966; Kasting et al. 1993), and they will not have seasons because their obliquities are eroded (Heller et al. 2011a,b). On moons, however, tides from the star are mostly negligible compared to the tidal drag from the planet. Thus, in most cases exomoons will be tidally locked to their host planet rather than to the star (Dole 1964; Gonzalez 2005; Henning et al. 2009; Kaltenegger 2010; Kipping et al. 2010) so that (*i.*) a satellite's rotation period will equal its orbital period about the planet, (*ii.*) a moon will orbit the planet in its equatorial plane (due to the Kozai mechanism and tidal evolution, Porter & Grundy 2011), and (*iii*.) a moon's rotation axis will be perpendicular to its orbit about the planet. A combination of (*ii*.) and (*iii*.) will cause the satellite to have the same obliquity with respect to the circumstellar orbit as the planet.

[1] A related but more anthropocentric circumstellar zone, termed "ecosphere", has been defined by Dole (1964, p. 64 therein). Whewell (1853, Chapter X, Section 4 therein) presented a more qualitative discussion of the so-called "Temperate Zone of the Solar System".

More massive planets are more resistive against the tidal erosion of their obliquities (Heller et al. 2011b); thus massive host planets of exomoons can maintain significant obliquities on timescales much larger than those for single terrestrial planets. Consequently, satellites of massive exoplanets could be located in the IHZ of low-mass stars while, firstly, their rotation would not be locked to their orbit around the star (but to the planet) and, secondly, they could experience seasons if the equator of their host planet is tilted against the orbital plane. Both aspects tend to reduce seasonal amplitudes of stellar irradiation (Cowan et al. 2012) and thereby stabilize exomoon climates.

An example is given by a potentially habitable moon in the Solar System, Titan. It is the only moon known to have a substantial atmosphere. Tides exerted by the Sun on Titan's host planet, Saturn, are relatively weak, which is why the planet could maintain its spin-orbit misalignment, or obliquity, $\psi_p$ of 26.7° (Norman 2011). Titan orbits Saturn in the planet's equatorial plane with no significant tilt of its rotation axis with respect to its circumplanetary orbit. Thus, the satellite shows a similar obliquity with respect to the Sun as Saturn and experiences strong seasonal modulations of insolation as a function of latitude, which leads to an alternation in the extents and localizations of its lakes and potential habitats (Wall et al. 2010). While tides from the Sun are negligible, Titan is tidally synchronized with Saturn (Lemmon et al. 1993) and has a rotation and an orbital period of ≈16d. Uranus, where $\psi_p \approx 97.9°$ (Harris & Ward 1982), illustrates that even more extreme scenarios are possible.

No exomoon has been detected so far, but it has been shown that exomoons with masses down to 20% the mass of Earth ($M_\oplus$) are detectable with the space-based *Kepler* telescope (Kipping et al. 2009). Combined measurements of a planet's transit timing variation (TTV) and transit duration variation (TDV) can provide information about the satellite's mass ($M_s$), its semi-major axis around the planet ($a_{ps}$) (Sartoretti & Schneider 1999; Simon et al. 2007; Kipping 2009a), and possibly about the inclination ($i$) of the satellite's orbit with respect to the orbit around the star (Kipping 2009b). Photometric transits of a moon in front of the star (Szabó et al. 2006; Lewis 2011; Kipping 2011a; Tusnski & Valio 2011), as well as mutual eclipses of a planet and its moon (Cabrera & Schneider 2007; Pál 2012), can provide information about its radius ($R_s$), and the photometric scatter peak analysis (Simon et al. 2012) can offer further evidence for the exomoon nature of candidate objects. Finally, spectroscopic investigations of the Rossiter-McLaughlin effect can yield information about the satellite's orbital geometry (Simon et al. 2010; Zhuang et al. 2012), although relevant effects require accuracies in stellar radial velocity of the order of a few centimeters per second (see also Kipping 2011a). Beyond, Peters & Turner (2013) suggest that direct imaging of extremely tidally heated exomoons will be possible with next-generation space telescopes. It was only recently that Kipping et al. (2012) initiated the first dedicated hunt for exomoons, based on *Kepler* observations. While we are waiting for their first discoveries, hints to exomoon-forming regions around planets have already been found (Mamajek et al. 2012).

In Section 2 of this paper, we consider general aspects of exomoon habitability to provide a basis for our work, while Section 3 is devoted to the description of the exomoon illumination plus tidal heating model. Section 4 presents a derivation of the critical orbit-averaged global flux and the description of habitable exomoon orbits, ultimately leading to the concept of the "habitable edge". In Section 5, we apply our model to putative exomoons around the first Neptune-sized[2] planet in the IHZ of a Sun-like star, Kepler-22b, and a much more massive, still to be confirmed planet candidate, the "Kepler Object of Interest" (KOI) 211.01[3], also supposed to orbit in the IHZ. We summarize our results with a discussion in Section 6. Detailed illustrations on how we derive our model are placed into the appendices.

## 2. Habitability of exomoons

So far, there have been only a few published investigations on exomoon habitability (Reynolds et al. 1987; Williams et al. 1997; Kaltenegger 2000; Scharf 2006; Porter & Grundy 2011). Other studies were mainly concerned with the observational aspects of exomoons (for a review see Kipping et al. 2012), their orbital stability (Barnes & O'Brien 2002; Domingos et al. 2006; Donnison 2010; Weidner & Horne 2010; Quarles et al. 2012; Sasaki et al. 2012), and eventually with the detection of biosignatures (Kaltenegger 2010). Thus, we provide here a brief overview of some important physical and biological aspects that must be accounted for when considering exomoon habitability.

---

[2] Planets with radii between $2R_\oplus$ and $6R_\oplus$ are designated Neptune-sized planets by the *Kepler* team (Batalha et al. 2012).

[3] Although KOI211.01 is merely a planet candidate we talk of it as a planet, for simplicity, but keep in mind its unconfirmed status.

Williams et al. (1997) were some of the first who proposed that habitable exomoons could be orbiting giant planets. At the time of their writing, only nine extrasolar planets, all of which are giant gaseous objects, were known. Although these bodies are not supposed to be habitable, Williams et al. argued that possible satellites of the jovian planets 16 Cygni B and 47 Ursae Majoris could offer habitats, because they orbit their respective host star at the outer edge of the habitable zone (Kasting et al. 1993). The main counter arguments against habitable exomoons were (*i.*) tidal locking of the moon with respect to the planet, (*ii.*) a volatile endowment of those moons, which would have formed in a circumplanetary disk, that is different from the abundances available for planets forming in a circumstellar disk, and (*iii.*) bombardment of high-energy ions and electrons within the magnetic fields of the jovian host planet and subsequent loss of the satellite's atmosphere. Moreover, (*iv.*) stellar forcing of a moon's upper atmosphere will constrain its habitability.

Point (*i.*), in fact, turns out as an advantage for Earth-sized satellites of giant planets over terrestrial planets in terms of habitability, by the following reasoning: Application of tidal theories shows that the rotation of extrasolar planets in the IHZ around low-mass stars will be synchronized on timescales $\ll$1Gyr (Dole 1964; Goldreich 1966; Kasting et al. 1993). This means one hemisphere of the planet will permanently face the star, while the other hemisphere will freeze in eternal darkness. Such planets might still be habitable (Joshi et al. 1997), but extreme weather conditions would strongly constrain the extent of habitable regions on the planetary surface (Heath & Doyle 2004; Spiegel et al. 2008; Heng & Vogt 2011; Edson et al. 2011; Wordsworth et al. 2011). However, considering an Earth-mass exomoon around a Jupiter-like host planet, within a few million years at most the satellite should be tidally locked to the planet – rather than to the star (Porter & Grundy 2011). This configuration would not only prevent a primordial atmosphere from evaporating on the illuminated side or freezing out on the dark side (*i.*) but might also sustain its internal dynamo (*iii.*). The synchronized rotation periods of putative Earth-mass exomoons around giant planets could be in the same range as the orbital periods of the Galilean moons around Jupiter (1.7d–16.7d) and as Titan's orbital period around Saturn ($\approx$16d) (NASA/JPL planetary satellite ephemerides)[4]. The longest possible length of a satellite's day compatible with Hill stability has been shown to be about $P_{*p}/9$, $P_{*p}$ being the planet's orbital period about the star (Kipping 2009a). Since the satellite's rotation period also depends on its orbital eccentricity around the planet and since the gravitational drag of further moons or a close host star could pump the satellite's eccentricity (Cassidy et al. 2009; Porter & Grundy 2011), exomoons might rotate even faster than their orbital period.

Finally, from what we know about the moons of the giant planets in the Solar System, the satellite's enrichment with volatiles (*ii.*) should not be a problem. Cometary bombardment has been proposed as a source for the dense atmosphere of Saturn's moon Titan, and it has been shown that even the currently atmosphere-free jovian moons Ganymede and Callisto should initially have been supplied with enough volatiles for an atmosphere (Griffith & Zahnle 1995). Moreover, as giant planets in the IHZ likely formed farther away from their star, that is, outside the snow line (Kennedy & Kenyon 2008), their satellites will be rich in liquid water and eventually be surrounded by substantial atmospheres.

The stability of a satellite's atmosphere (*iv.*) will critically depend on its composition, the intensity of stellar extreme ultraviolet radiation (EUV), and the moon's surface gravity. Nitrogen-dominated atmospheres may be stripped away by ionizing EUV radiation, which is a critical issue to consider for young (Lichtenberger et al. 2010) and late-type (Lammer et al. 2009) stars. Intense EUV flux could heat and expand a moon's upper atmosphere so that it can thermally escape due to highly energetic radiation (*iii.*), and if the atmosphere is thermally expanded beyond the satellite's magnetosphere, then the surrounding plasma may strip away the atmosphere nonthermally. If Titan were to be moved from its roughly 10AU orbit around the Sun to a distance of 1AU (AU being an astronomical unit, i.e., the average distance between the Sun and Earth), then it would receive about 100 times more EUV radiation, leading to a rapid loss of its atmosphere due to the moon's smaller mass, compared to Earth. For an Earth-mass moon at 1AU from the Sun, EUV radiation would need to be less than 7 times the Sun's present-day EUV emission to allow for a long-term stability of a nitrogen atmosphere. $CO_2$ provides substantial cooling of an atmosphere by infrared radiation, thereby counteracting thermal expansion and protecting an atmosphere's nitrogen inventory (Tian 2009).

A minimum mass of an exomoon is required to drive a magnetic shield on a billion-year timescale ($M_s \gtrsim 0.1 M_\oplus$, Tachinami et al. 2011); to sustain a substantial, long-lived atmosphere ($M_s \gtrsim 0.12 M_\oplus$, Williams et al. 1997; Kaltenegger 2000); and to drive tectonic activity ($M_s \gtrsim 0.23 M_\oplus$, Williams et al. 1997), which is necessary to maintain plate tectonics and to support the carbon-silicate cycle. Weak internal dynamos have been detected in Mercury and Ganymede (Kivelson et al. 1996; Gurnett et al. 1996), suggesting that satellite masses $> 0.25 M_\oplus$ will be adequate for considerations of exomoon habitability. This lower limit, however, is not a fixed number. Further sources of energy – such as radiogenic and tidal heating, and the effect of a moon's composition and structure – can alter our limit in either direction. An upper mass limit is given by the fact that increasing mass leads to high pressures in the moon's interior, which will increase the mantle viscosity and depress heat transfer throughout the mantle as well as in the core. Above a critical mass, the dynamo is strongly suppressed and becomes too weak to generate a magnetic field or sustain plate tectonics. This maximum mass can be placed around $2M_{\oplus}$ (Gaidos et al. 2010; Noack & Breuer 2011; Stamenković et al. 2011). Summing up these conditions, we expect approximately Earth-mass moons to be habitable, and these objects could be detectable with the newly started *Hunt for Exomoons with Kepler* (HEK) project (Kipping et al. 2012).

[4] Maintained by Robert Jacobson, http://ssd.jpl.nasa.gov.

### 2.1 Formation of massive satellites

The largest and most massive moon in the Solar System, Ganymede, has a radius of only $\approx 0.4R_{\oplus}$ ($R_{\oplus}$ being the radius of Earth) and a mass of $\approx 0.025M_{\oplus}$. The question as to whether much more massive moons could have formed around extrasolar planets is an active area of research. Canup & Ward (2006) have shown that moons formed in the circumplanetary disk of giant planets have masses $\lesssim 10^{-4}$ times that of the planet's mass. Assuming satellites formed around Kepler-22b, their masses will thus be $2.5 \times 10^{-3} M_{\oplus}$ at most, and around KOI211.01 they will still weigh less than Earth's Moon. Mass-constrained in situ formation becomes critical for exomoons around planets in the IHZ of low-mass stars because of the observational lack of such giant planets. An excellent study on the formation of the Jupiter and the Saturn satellite systems is given by Sasaki et al. (2010), who showed that moons of sizes similar to Io, Europa, Ganymede, Callisto, and Titan should build up around most gas giants. What is more, according to their Fig. 5 and private communication with Takanori Sasaki, formation of Mars- or even Earth-mass moons around giant planets is possible. Depending on whether or not a planet accretes enough mass to open up a gap in the protostellar disk, these satellite systems will likely be multiple and resonant (as in the case of Jupiter), or contain only one major moon (see Saturn). Ogihara & Ida (2012) extended these studies to explain the compositional gradient of the jovian satellites. Their results explain why moons rich in water are farther away from their giant host planet and imply that capture in 2:1 orbital resonances should be common.

Ways to circumvent the impasse of insufficient satellite mass are the gravitational capture of massive moons (Debes & Sigurdsson 2007; Porter & Grundy 2011; Quarles et al. 2012), which seems to have worked for Triton around Neptune (Goldreich et al. 1989; Agnor & Hamilton 2006); the capture of Trojans (Eberle et al. 2011); gas drag in primordial circumplanetary envelopes (Pollack et al. 1979); pull-down capture trapping temporary satellites or bodies near the Lagrangian points into stable orbits (Heppenheimer & Porco 1977; Jewitt & Haghighipour 2007); the coalescence of moons (Mosqueira & Estrada 2003); and impacts on terrestrial planets (Canup 2004; Withers & Barnes 2010; Elser et al. 2011). Such moons would correspond to the irregular satellites in the Solar System, as opposed to regular satellites that form in situ. Irregular satellites often follow distant, inclined, and often eccentric or even retrograde orbits about their planet (Carruba et al. 2002). For now, we assume that Earth-mass extrasolar moons – be they regular or irregular – exist.

### 2.2 Deflection of harmful radiation

A prominent argument against the habitability of moons involves high-energy particles, which a satellite meets in the planet's radiation belt. Firstly, this ionizing radiation could strip away a moon's atmosphere, and secondly it could avoid the buildup of complex molecules on its surface. In general, the process in which incident particles lose part of their energy to a planetary atmosphere or surface to excite the target atoms and molecules is called sputtering. The main sources for sputtering on Jupiter's satellites are the energetic, heavy ions $O^+$ and $S^+$, as well as $H^+$, which give rise to a steady flux of $H_2O$, OH, $O_2$, $H_2$, O, and H from Ganymede's surface (Marconi 2007). A moon therefore requires a substantial magnetic field that is strong enough to embed the satellite in a protective bubble inside the planet's powerful magnetosphere. The only satellite in the Solar System with a substantial magnetic shield of roughly 750nT is Ganymede (Kivelson et al. 1996). The origin of this field is still subject to debate because it can only be explained by a very specific set of initial and compositional configurations (Bland et al. 2008), assuming that it is generated in the moon's core.

For terrestrial planets, various models for the strength of global dipolar magnetic fields $B_{\rm dip}$ as a function of planetary mass and rotation rate exist, but none has proven exclusively valid. Simulations of planetary thermal evolution have shown that $B_{\rm dip}$ increases with mass (Tachinami et al. 2011; Zuluaga & Cuartas 2012) and rotation frequency (Lopez-Morales et al. 2012). The spin of exomoons will be determined by tides from the planet, and rotation of an Earth-sized exomoon in the IHZ can be much faster than rotation of an Earth-sized planet orbiting a star. Thus, an exomoon could be prevented from tidal synchronization with the host star – in support of an internal dynamo and thus magnetic shielding against energetic irradiation from the planet and the star. Some studies suggest that even extremely slow rotation would allow for substantial magnetic shielding, provided convection in the planet's or moon's mantle is strong enough (Olson & Christensen 2006). In this case, tidal locking would not be an issue for magnetic shielding.

The picture of magnetic shielding gets even more complicated when tidal heating is considered, which again depends on the orbital parameters. In the Moon, tidal heating, mostly induced by the Moon's obliquity of 6.68° against its orbit around Earth, occurs dominantly in the core (Kaula 1964; Peale & Cassen 1978). On Io, however, where tidal heating stems from Jupiter's effect on the satellite's eccentricity, dissipation occurs mostly in the mantle (Segatz et al. 1988). In the former case, tidal heating might enhance the temperature gradient between the core and the mantle and thereby also enhance convection and finally the strength of the magnetic shielding; in the latter case, tidal heating might decrease convection. Of course, the magnetic properties of terrestrial worlds will evolve and, when combined with the evolution of EUV radiation and stellar wind from the host star, define a time-dependent magnetically restricted habitable zone (Khodachenko et al. 2007; Zuluaga et al. 2012).

We conclude that radiation of highly energetic particles does not ultimately preclude exomoon habitability. In view of possible deflection due to magnetic fields on a massive satellites, it is still reasonable to consider the habitability of exomoons.

### 2.3 Runaway greenhouse

On Earth, the thermal equilibrium temperature of incoming and outgoing radiation is 255K. However, the mean surface temperature is 289K. The additional heating is driven by the greenhouse effect (Kasting 1988), which is a crucial phenomenon to the habitability of terrestrial bodies. The strength of the greenhouse effect depends on numerous variables – most importantly on the inventory of greenhouse gases, the albedo effect of clouds, the amount of liquid surface water, and the spectral energy distribution of the host star.

Simulations have shown that, as the globally absorbed irradiation on a water-rich planetary body increases, the atmosphere gets enriched in water vapor until it gets opaque. For an Earth-like body, this imposes a limit of about 300W/m$^2$ to the thermal radiation that can be emitted to space. If the global flux exceeds this limit, the body is said to be a runaway greenhouse. Water vapor can then leave the troposphere through the tropopause and reach the stratosphere, where photodissociation by stellar UV radiation allows the hydrogen to escape to space, thereby desiccating the planetary body. While boiling oceans, high surface temperatures, or high pressures *can* make a satellite uninhabitable, water loss *does* by definition. Hence, we will use the criterion of a runaway greenhouse to define an exomoon's habitability.

Surface temperatures strongly depend on the inventory of greenhouse gases, for example, $CO_2$. The critical energy flux $F_{RG}$ for a runaway greenhouse, however, does not (Kasting 1988; Goldblatt & Watson 2012). As in Barnes et al. (2013), who discussed how the interplay of stellar irradiation and tidal heating can trigger a runaway greenhouse on exoplanets, we will use the semi-analytical approach of Pierrehumbert (2010) for the computation of $F_{RG}$:

$$F_{\rm RG} = o\,\sigma_{\rm SB}\left(\frac{l}{R\ln\left(P'/\sqrt{\frac{2P_0 g_{\rm s}(M_{\rm s},R_{\rm s})}{k_0}}\right)}\right)^4 \qquad (1)$$

with

$$P' = P_{\rm ref}\exp\left\{\frac{l}{R\,T_{\rm ref}}\right\}\ , \qquad (2)$$

$P_{\rm ref}$ = 610.616Pa, $l$ is the latent heat capacity of water, $R$ is the universal gas constant, $T_{\rm ref}$ = 273.13K, $o$ = 0.7344 is a constant designed to match radiative transfer simulations, $\sigma_{\rm SB}$ is the Stefan-Boltzmann constant, $P_0$ = 104Pa is the pressure at which the absorption line strengths of water vapor are evaluated, $g_{\rm s} = GM_{\rm s}/R_{\rm s}^2$ is the gravitational acceleration at the satellite's surface, and $k_0$ = 0.055 is the gray absorption coefficient at standard temperature and pressure. Recall that the runaway greenhouse does not depend on the composition of the atmosphere, other than it contains water. As habitability requires water and Eq. (1) defines a limit above which the satellite will lose it, the formula provides a conservative limit to habitability.

In addition to the maximum flux $F_{RG}$ to allow for a moon to be habitable, one may think of a minimum flux required to prevent the surface water from freezing. On terrestrial exoplanets, this freezing defines the outer limit of the stellar IHZ. On exomoons, the extra light from the planetary reflection and thermal emission as well as tidal heating in the moon will move the circumstellar habitable zone away from the star, whereas eclipses will somewhat counterbalance this effect. While it is clear that a moon under strong tidal heating will not be habitable, it is not clear to what extent it might actually support habitability (Jackson et al. 2008). Even a relatively small tidal heating flux of a few watts per square meter could render an exomoon inhospitable; see Io's global volcanism, where tidal heating is a mere $2\mathrm{W/m^2}$ (Spencer et al. 2000). Without applying sophisticated models for the moon's tidal heating, we must stick to the irradiation aspect to define an exomoon's circumstellar habitable zone. At the outer edge of the stellar IHZ, the host planet will be cool and reflected little stellar flux. Neglecting tidal heating as well thermal emission and reflection from the planet, the minimum flux for an Earth-like moon to be habitable will thus be similar to that of an Earth-like planet at the same orbital distance to the star. Below, we will only use the upper flux limit from Eq. (1) to constrain the orbits of habitable exomoons. This will lead us to the concept of the circumplanetary "habitable edge".

## 3. Energy reservoirs on exomoons

Life needs liquid water and energy, but an oversupply of energy can push a planet or an exomoon into a runaway greenhouse and thereby make it uninhabitable. The critical, orbit-averaged energy flux for an exomoon to turn into a runaway greenhouse is around $300\mathrm{W/m^2}$, depending on the moon's surface gravity and atmospheric composition (Kasting 1988; Kasting et al. 1993; Selsis et al. 2007; Pierrehumbert 2010; Goldblatt & Watson 2012). An exomoon will thus only be habitable in a certain range of summed irradiation and tidal heat flux (Barnes et al. 2013).

We consider four energy reservoirs and set them into context with the IHZ: (*i.*) stellar illumination, (*ii.*) stellar reflected light from the planet, (*iii.*) thermal radiation from the planet, and (*iv.*) tidal heating on the moon. Here, primordial heat from the moon's formation and radiogenic decay is neglected, and it is assumed that the moon's rotation is tidally locked to its host planet, as is the case for almost all the moons in the Solar System. Our irradiation model includes arbitrary orbital eccentricities $e_{*\mathrm{p}}$ of the planet around the star[5]. While we compute tidal heating on the satellite as a function of its orbital eccentricity $e_{\mathrm{ps}}$ around the planet, we assume $e_{\mathrm{ps}} = 0$ in the parametrization of the moon's irradiation. This is appropriate because typically $e_{\mathrm{ps}} \ll 0.1$. Bolmont et al. (2011) studied the tidal evolution of Earth-mass objects around brown dwarfs, a problem which is similar to an Earth-mass moon orbiting a jovian planet, and found that tidal equilibrium occurs on very short timescales compared to the lifetime of the system. For non-zero eccentricities, ($e_{\mathrm{ps}} \neq 0$), the moon will not be tidally locked. But since $e_{\mathrm{ps}} \ll 0.1$, the moon's rotation will librate around an equilibrium orientation toward the planet, and the orbital mean motion will still be almost equal to the rotational mean motion (for a review on Titan's libration, see Sohl et al. 1995). By reasons specified by Heller et al. (2011b), we also assume that the obliquity of the satellite with respect to its orbit around the planet has been eroded to 0°, but we allow for arbitrary inclinations $i$ of the moon's orbit with respect to the orbit of the planet-moon barycenter around the star. If one assumed that the moon always orbits above the planet's equator, that would imply that $i$ is equal to the planetary obliquity $\psi_{\mathrm{p}}$, which is measured with respect to the planet's orbit around the star. We do not need this assumption for the derivation of our equations, but since $\psi_{\mathrm{p}} \approx i$ for all the large moons in the Solar System, except Triton, observations or numerical predictions of $\psi_{\mathrm{p}}$ (Heller et al. 2011b) can provide reasonable assumptions for $i$.

In our simulations, we consider two prototype moons: one rocky Earth-mass satellite with a rock-to-mass fraction of 68% (similar to Earth) and one water-rich satellite with the tenfold mass of Ganymede and an ice-to-mass fraction of 25% (Fortney et al. 2007). The remaining constituents are assumed to be iron for the Earth-mass moon and silicates for the Super-Ganymede. The more massive and relatively dry moon represents what we guess a captured, Earth-like exomoon could be like, while the latter one corresponds to a satellite that has formed in situ. Note that a mass of $10M_{\mathrm{G}}$ ($M_{\mathrm{G}}$ being the mass of Ganymede) corresponds to roughly $0.25M_{\oplus}$, which is slightly above the detection limit for combined TTV and TDV with *Kepler* (Kipping et al. 2009). Our assumptions for the Super-Ganymede composition are backed up by observations of the Jupiter and Saturn satellite systems (Showman & Malhotra 1999; Grasset et al. 2000) as well as terrestrial planet and satellite formation studies (Kuchner 2003; Ogihara & Ida 2012). These papers show that in situ formation naturally generates water-rich moons and that such objects can retain their water reservoir for billions of years against steady hydrodynamic escape. Concerning the habitability of the water-rich Super-Ganymede, we do not rely on any assumptions concerning possible life forms in such water worlds. Except for the possible strong heating in a water-rich atmosphere (Matsui & Abe 1986; Kuchner 2003), we see no reason why ocean moons should not be hospitable, in particular against the background that life on Earth arose in (possibly hot) oceans or freshwater seas.

[5] In the following, a parameter index "$*$" will refer to the star, "p" to the planet, and "s" to the satellite. The combinations "$*$p" and "ps", e.g., for the orbital eccentricities $e_{*\mathrm{p}}$ and $e_{\mathrm{ps}}$, refer to systems of a star plus a planet and a planet plus a satellite, respectively. For a vector, e.g. $\vec{r}_{\mathrm{p}*}$, the first letter indicates the starting point (in this case the planet) and the second index locates the endpoint (here the star).

For the sake of consistency, we derive the satellites' radii $R_s$ from planetary structure models (Fortney et al. 2007). In the case of the Earth-mass moon, we obtain $R_s = 1R_\oplus$, and for the much lighter but water-dominated Super-Ganymede $R_s = 0.807R_\oplus$. Equation (1) yields a critical flux of 295W/m$^2$ for the Earth-mass moon, and 266W/m$^2$ for the Super-Ganymede satellite. The bond albedo of both moons is assumed to be 0.3, similar to the mean albedo of Earth and of the Galilean satellites (Clark 1980). In the following, we call our Earth-like and Super-Ganymede satellites our "prototype moons". Based on the summary of observations and the model for giant planet atmospheres provided by Madhusudhan & Burrows (2012), we also use a bond albedo of 0.3 for the host planet, although higher values might be reasonable due to the formation of water clouds at distances 1AU from the host star (Burrows et al. 2006a). Mass and radius of the planet are not fixed in our model, but we will mostly refer to Jupiter-sized host planets.

### 3.1. Illumination

The total bolometric illumination on a moon is given by the stellar flux ($f_*$), the reflection of the stellar light from the planet ($f_r$), and the planetary thermal emission ($f_t$). Their variation will be a function of the satellite's orbital phase $0 \leq \varphi_{ps}(t) = (t{-}t_0)/P_{ps} \leq 1$ around the planet (with $t$ being time, $t_0$ as the starting time [0 in our simulations], and $P_{ps}$ as the period of the planet-moon orbit), the orbital phase of the planet-moon duet around the star ($\varphi_{*p}$, which is equivalent to the mean anomaly $\mathfrak{M}_{*p}$ divided by $2\pi$), and will depend on the eccentricity of the planet around the star ($e_{*p}$), on the inclination ($i$) of the two orbits, on the orientation of the periapses ($\eta$), as well as on longitude and latitude on the moon's surface ($\phi$ and $\theta$).

In Fig. 1, we show the variation of the satellite's illumination as a function of the satellite's orbital phase $\varphi_{ps}$. For this plot, the orbital phase of the planet-moon pair around the star $\varphi_{*p} = 0$ and $i = 0$. Projection effects due to latitudinal variation have been neglected, starlight is assumed to be plane-parallel, and radii and distances are not to scale.

In our irradiation model of a tidally locked satellite, we neglect clouds, radiative transfer, atmospheric circulation, geothermal flux[6], thermal inertia, and so on, and we make use of four simplifications:

($i$.) We assume the planet casts no penumbra on the moon. There is either total illumination from the star or none. This assumption is appropriate since we are primarily interested in the key contributions to the moon's climate.

($ii$.) The planet is assumed to be much more massive than the moon, and the barycenter of the planet-moon binary is placed at the center of the planet. Even if the planet and the moon had equal masses, corrections would be small since the range between the planet-moon barycenter and the star $\gg a_{ps}$.

($iii$.) For the computation of the irradiation, we treat the moon's orbit around the planet as a circle. The small eccentricities which we will consider later for tidal heating will not modify our results significantly.

($iv$.) The distance between the planet-satellite binary and the star does not change significantly over one satellite orbit, which is granted when either $e_{*p}$ is small or $P_{ps} \ll P_{*p}$.

In the following, we present the general results of our mathematical derivation. For a more thorough description and discussions of some simple cases, see Appendices A and B.

#### 3.1.1 Illumination from the star

The stellar flux on the substellar point on the moon's surface will have a magnitude $L_*/(4\pi r_{s*}(t)^2)$, where $L_*$ is stellar luminosity and $\vec{r}_{s*}$ is the vector from the satellite to the star. We multiply this quantity with the surface normal $\vec{n}_{\phi,\theta}/n_{\phi,\theta}$ on the moon and $\vec{r}_{s*}/r_{s*}$ to include projection effects on a location ($\phi$,$\theta$). This yields

$$f_*(t) = \frac{L_*}{4\pi \vec{r}_{s*}(t)^2} \frac{\vec{r}_{s*}(t)}{r_{s*}(t)} \frac{\vec{n}_{\phi,\theta}(t)}{n_{\phi,\theta}(t)} \ . \qquad (3)$$

[6] Tidal heating will be included below, but we will neglect geothermal feedback between tidal heating and irradiation.

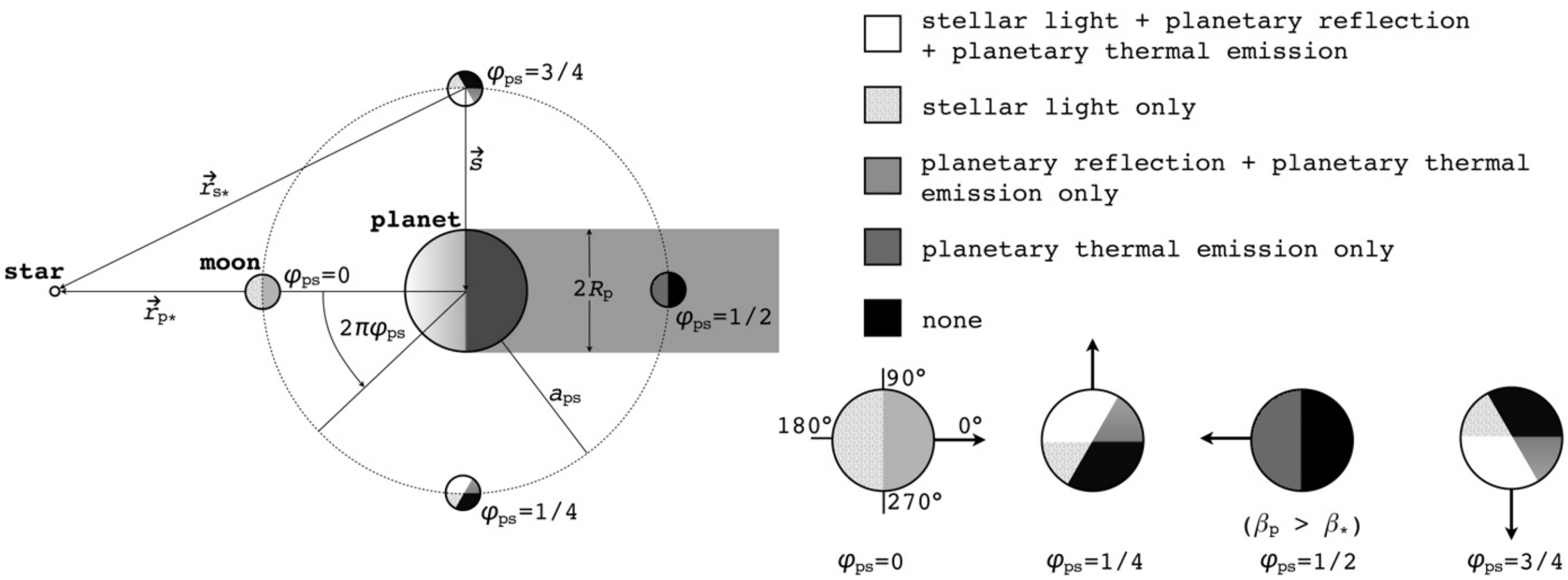

**FIG. 1.** Geometry of the triple system of a star, a planet, and a moon with illuminations indicated by different shadings (pole view). For ease of visualization, the moon's orbit is coplanar with the planet's orbit about the star and the planet's orbital position with respect to the star is fixed. Combined stellar and planetary irradiation on the moon is shown for four orbital phases. Projection effects as a function of longitude $\phi$ and latitude $\theta$ are ignored, and we neglect effects of a penumbra. Radii and distances are not to scale, and starlight is assumed to be plane-parallel. In the right panel, the surface normal on the subplanetary point is indicated by an arrow. For a tidally locked moon this spot is a fixed point on the moon's surface. For $\varphi_{ps} = 0$ four longitudes are indicated.

If $\vec{r}_{s*}$ and $\vec{n}$ have an antiparallel part, then $f_* < 0$, which is meaningless in our context, and we set $f_*$ to zero. The task is now to find $\vec{r}_{s*}(t)$ and $\vec{n}_{\phi,\theta}(t)$. Therefore, we introduce the surface vector from the subplanetary point on the satellite to the planet, $\vec{s} \equiv \vec{n}_{0,0}$, and the vector from the planet to the star, $\vec{r}_{p*}(t)$, which gives $\vec{r}_{s*}(t) = \vec{r}_{p*}(t) + \vec{s}(t)$ (see Fig. 1). Applying Kepler's equations of motion, we deduce $\vec{r}_{p*}(t)$; and with a few geometric operations (see Appendix A) we obtain $\vec{n}_{\phi,\theta}(t)$:

$$\vec{r}_{p*}(t) = -a_{*p} \begin{pmatrix} \tilde{c} - e_{*p} \\ \sqrt{1-e_{*p}^2}\,\tilde{s} \\ 0 \end{pmatrix} \quad (4) \qquad \vec{n}_{\phi,\theta}(t) = a_{ps} \begin{pmatrix} -\bar{s}S\tilde{C} + \bar{c}(\tilde{C}cC - \tilde{S}s) \\ -\bar{s}S\tilde{S} + \bar{c}(\tilde{S}cC - \tilde{C}s) \\ \bar{s}C + \bar{c}cS \end{pmatrix} \quad (5)$$

with

$$\begin{aligned} c &= \cos\left(2\pi(\varphi_{ps}(t) + \frac{\phi}{360^\circ})\right) & s &= \sin\left(2\pi(\varphi_{ps}(t) + \frac{\phi}{360^\circ})\right) \\ \bar{c} &= \cos\left(\theta\frac{\pi}{180^\circ})\right) & \bar{s} &= \sin\left(\theta\frac{\pi}{180^\circ})\right) \\ \tilde{c} &= \cos\left(E_{*p}(t)\right) & \tilde{s} &= \sin\left(E_{*p}(t)\right) \\ C &= \cos\left(i\frac{\pi}{180^\circ}\right) & S &= \sin\left(i\frac{\pi}{180^\circ}\right) \\ \tilde{C} &= \cos\left(\eta\frac{\pi}{180^\circ}\right) & \tilde{S} &= \sin\left(\eta\frac{\pi}{180^\circ}\right) \end{aligned} \quad (6)$$

where $i$, $\eta$, $0 \leq \phi \leq 360°$, and $0 \leq \theta \leq 90°$ are provided in degrees, and $\phi$ and $\theta$ are measured from the subplanetary point (see Fig. 1).

$$E_{*p}(t) - e_{*p}\sin\left(E_{*p}(t)\right) = \mathfrak{M}_{*p}(t) \quad (7)$$

defines the eccentric anomaly $E_{*p}$ and

$$\mathfrak{M}_{*p}(t) = 2\pi\frac{(t-t_0)}{P_{*p}} \quad (8)$$

is the mean anomaly. The angle $\eta$ is the orientation of the lowest point of the moon's inclined orbit with respect to the star at periastron (see Appendix A). *Kepler's equation* (Eq. 7) is a transcendental function which we solve numerically.

To compute the stellar flux over one revolution of the moon around the planet, we put the planet-moon duet at numerous orbital phases around the star (using a fixed time step d$t$), thus $\tilde{c}$ and $\tilde{s}$ will be given. At each of these positions, we then evolve $\varphi_{\rm ps}$ from 0 to 1. With this parametrization, the moon's orbit around the planet will always start at the left, corresponding to Fig. 1, and it will be more facile to interpret the phase functions. If we were to evolve the moon's orbit consistently, $2\pi\mathfrak{M}_{*\rm p}$ would have to be added to the arguments of $c$ and $s$. Our simplification is appropriate as long as $\vec{r}_{*\rm p}$ does not change significantly over one satellite orbit. Depending on the orientation of an eventual inclination between the two orbits and depending on the orbital position of the planet-moon system around the star, the can be eclipsed by the planet for a certain fraction of $\varphi_{\rm ps}$ as seen from the moon. This phenomenon might have significant impacts on exomoon climates. Eclipses occur if the perpendicular part

$$r_\perp = \sin\left( \arccos\left( \frac{\vec{r}_{\rm s*}\,\vec{r}_{\rm p*}}{|\vec{r}_{\rm s*}||\vec{r}_{\rm p*}|} \right) \right) |\vec{r}_{\rm s*}| \qquad (9)$$

of $\vec{r}_{\rm s*}$ with respect to $\vec{r}_{\rm p*}$ is smaller than the radius of the planet and if $|\vec{r}_{\rm s*}| > |\vec{r}_{\rm p*}|$, that is, if the moon is behind the planet as seen from the star and not in front of it. The angular diameters of the star and the planet, $\beta_*$ and $\beta_{\rm p}$, respectively, are given by

$$\begin{aligned} \beta_* &= 2\arctan\left( \frac{R_*}{a_{*\rm p} + a_{\rm ps}} \right) \\ \beta_{\rm p} &= 2\arctan\left( \frac{R_{\rm p}}{a_{\rm ps}} \right) \ . \end{aligned} \qquad (10)$$

If $\beta_{\rm p} > \beta_*$ then the eclipse will be total. Otherwise the stellar flux will be diminished by a factor $[1 - (\beta_{\rm p}/\beta_*)^2]$.

### 3.1.2 Illumination from the planet

We now consider two contributions to exomoon illumination from the planet, namely, reflection of stellar light ($f_{\rm r}$) and thermal radiation ($f_{\rm t}$). If the planet's rotation period is ≲ 1d, then the stellar irradiation will be distributed somewhat smoothly over longitude. However, for a planet which is tidally locked to the star, the illuminated hemisphere will be significantly warmer than the back side. In our model, the bright side of the planet has a temperature $T^{\rm b}_{\rm eff,p}$, and the dark back side has a temperature $T^{\rm d}_{\rm eff,p}$ (see Appendix B). With ${\rm d}T = T^{\rm b}_{\rm eff,p} - T^{\rm d}_{\rm eff,p}$ as the temperature difference between the hemispheres and $\alpha_{\rm p}$ as the planet's bond albedo, that is, the fraction of power at all wavelengths scattered back into space, thermal equilibrium yields

$$p(T^{\rm b}_{\rm eff,p}) \equiv (T^{\rm b}_{\rm eff,p})^4 + (T^{\rm b}_{\rm eff,p} - {\rm d}T)^4 - T^4_{\rm eff,*} \frac{(1-\alpha_{\rm p})R_*^2}{2\vec{r}^{\,2}_{*\rm p}} = 0 \ . \qquad (11)$$

For a given d$T$, we search for the zero points of the polynomial $p(T^{\rm b}_{\rm eff,p})$ numerically. In our prototype system at 1AU from a Sun-like star and choosing d$T$ = 100K, Eq. (11) yields $T^{\rm b}_{\rm eff,p} = 291\,{\rm K}$ and $T^{\rm d}_{\rm eff,p} = 191\,{\rm K}$. Finally, the thermal flux received by the moon from the planet turns out as

$$f_{\rm t}(t) = \frac{R_{\rm p}^2 \sigma_{\rm SB}}{a_{\rm ps}^2} \cos\left( \frac{\phi\pi}{180^\circ} \right) \cos\left( \frac{\theta\pi}{180^\circ} \right) \times \left[ (T^{\rm b}_{\rm eff,p})^4 \xi(t) + (T^{\rm d}_{\rm eff,p})^4 \left(1 - \xi(t)\right) \right] \ , \qquad (12)$$

where

$$\xi(t) = \frac{1}{2} \left\{ 1 + \cos\Big(\vartheta(t)\Big) \cos\Big(\Phi(t) - \nu_{*\rm p}(t)\Big) \right\} \qquad (13)$$

weighs the contributions from the two hemispheres,

$$\nu_{*\mathrm{p}}(t) = \arccos\left(\frac{\cos(E_{*\mathrm{p}}(t)) - e_{*\mathrm{p}}}{1 - e_{*\mathrm{p}}\cos(E_{*\mathrm{p}}(t))}\right) \qquad (14)$$

is the true anomaly,

$$\Phi(t) = 2\arctan\left(\frac{s_{\mathrm{y}}(t)}{\sqrt{s_{\mathrm{x}}^2(t) + s_{\mathrm{y}}^2(t)} + s_{\mathrm{x}}(t)}\right)$$

$$\vartheta(t) = \frac{\pi}{2} - \arccos\left(\frac{s_{\mathrm{y}}(t)}{\sqrt{s_{\mathrm{x}}^2(t) + s_{\mathrm{y}}^2(t) + s_{\mathrm{z}}^2(t)}}\right) \ , \qquad (15)$$

and $s_{\mathrm{x,y,z}}$ are the components of $\vec{s} = (s_{\mathrm{x}}, s_{\mathrm{y}}, s_{\mathrm{z}})$ (see Appendix B).

Additionally, the planet reflects a portion $\pi R_{\mathrm{p}}^2 \alpha_{\mathrm{p}}$ of the incoming stellar light. Neglecting that the moon blocks a small fraction of the starlight when it passes between the planet and the star (< 1% for an Earth-sized satellite around a Jupiter-sized planet), we find that the moon receives a stellar flux

$$f_{\mathrm{r}}(t) = \frac{R_*^2 \sigma_{\mathrm{SB}} T_{\mathrm{eff},*}^4}{r_{\mathrm{p}*}^2} \frac{\pi R_{\mathrm{p}}^2 \alpha_{\mathrm{p}}}{a_{\mathrm{ps}}^2} \cos\left(\frac{\phi\pi}{180°}\right) \cos\left(\frac{\theta\pi}{180°}\right) \xi(t) \qquad (16)$$

from the planet.

In Fig. 2, we show how the amplitudes of $f_{\mathrm{t}}(t)$ and $f_{\mathrm{r}}(t)$ compare. Therefore, we neglect the time dependence and compute simply the maximum possible irradiation on the moon's subplanetary point as a function of the moon's orbit around the planet, which occurs in our model when the moon is over the substellar point of the planet. Then it receives maximum reflection and thermal flux at the same time. For our prototype system, it turns out that $f_{\mathrm{t}} > f_{\mathrm{r}}$ at a given planet-moon distance only if the planet has an albedo $\lesssim 0.1$, which means that it needs to be almost black. The exact value, $\alpha_{\mathrm{p}} = 0.093$ in this case, can be obtained by comparing Eqs. (12) and (16) (see Appendix B). For increasing $\alpha_{\mathrm{p}}$, stellar reflected flux dominates more and more; and for $\alpha_{\mathrm{p}} \gtrsim 0.6$, $f_*$ is over a magnitude stronger than $f_{\mathrm{t}}$.

The shapes of the curves can be understood intuitively, if one imagines that at a fixed semi-major axis (abscissa) the reflected flux received on the moon increases with increasing albedo (ordinate), whereas the planet's thermal flux increases when it absorbs more stellar light, which happens for decreasing albedo.

The shaded area in the upper left corner of the figure indicates where the sum of maximum $f_{\mathrm{t}}$ and $f_{\mathrm{r}}$ exceeds the limit of 295W/m$^2$ for a runaway greenhouse on an Earth-sized moon. Yet a satellite in this part of the parameter space would not necessarily be uninhabitable, because firstly it would only be subject to intense planetary radiation for less than about half its orbit, and secondly eclipses could cool the satellite half an orbit later. Moons at $a_{\mathrm{ps}} \lesssim 4R_{\mathrm{p}}$ are very likely to experience eclipses. Note that a moon's orbital eccentricity $e_{\mathrm{ps}}$ will have to be almost perfectly zero to avoid intense tidal heating in such close orbits (see Section 3.2).

Since we use $a_{\mathrm{ps}}$ in units of planetary radii, $f_{\mathrm{t}}$ and $f_{\mathrm{r}}$ are independent of $R_{\mathrm{p}}$. We also show a few examples of Solar System moons, where we adopted 0.343 for Jupiter's bond albedo (Hanel et al. 1981), 0.342 for Saturn (Hanel et al. 1983), 0.32 for Uranus (Neff et al. 1985; Pollack et al. 1986; Pearl et al. 1990), 0.29 for Neptune (Neff et al. 1985; Pollack et al. 1986; Pearl & Conrath 1991), and 0.3 for Earth. Flux contours are not directly applicable to the indicated moons because the host planets Jupiter, Saturn, Uranus, and Neptune do not orbit the Sun at 1AU, as assumed for our prototype exomoon system[7]. Only the position of Earth's moon, which receives a maximum of 0.35W/m$^2$ reflected light from Earth, reproduces the true Solar System values. The Roche radius for a fluid-like body (Weidner & Horne 2010, and references therein) is indicated with a gray line at $2.07R_{\mathrm{p}}$.

[7] At a distance of 5.2AU from the Sun, Europa receives roughly 0.5W/m$^2$ reflected light from Jupiter, when it passes the planet's subsolar point. Jupiter's thermal flux on Europa is negligible.

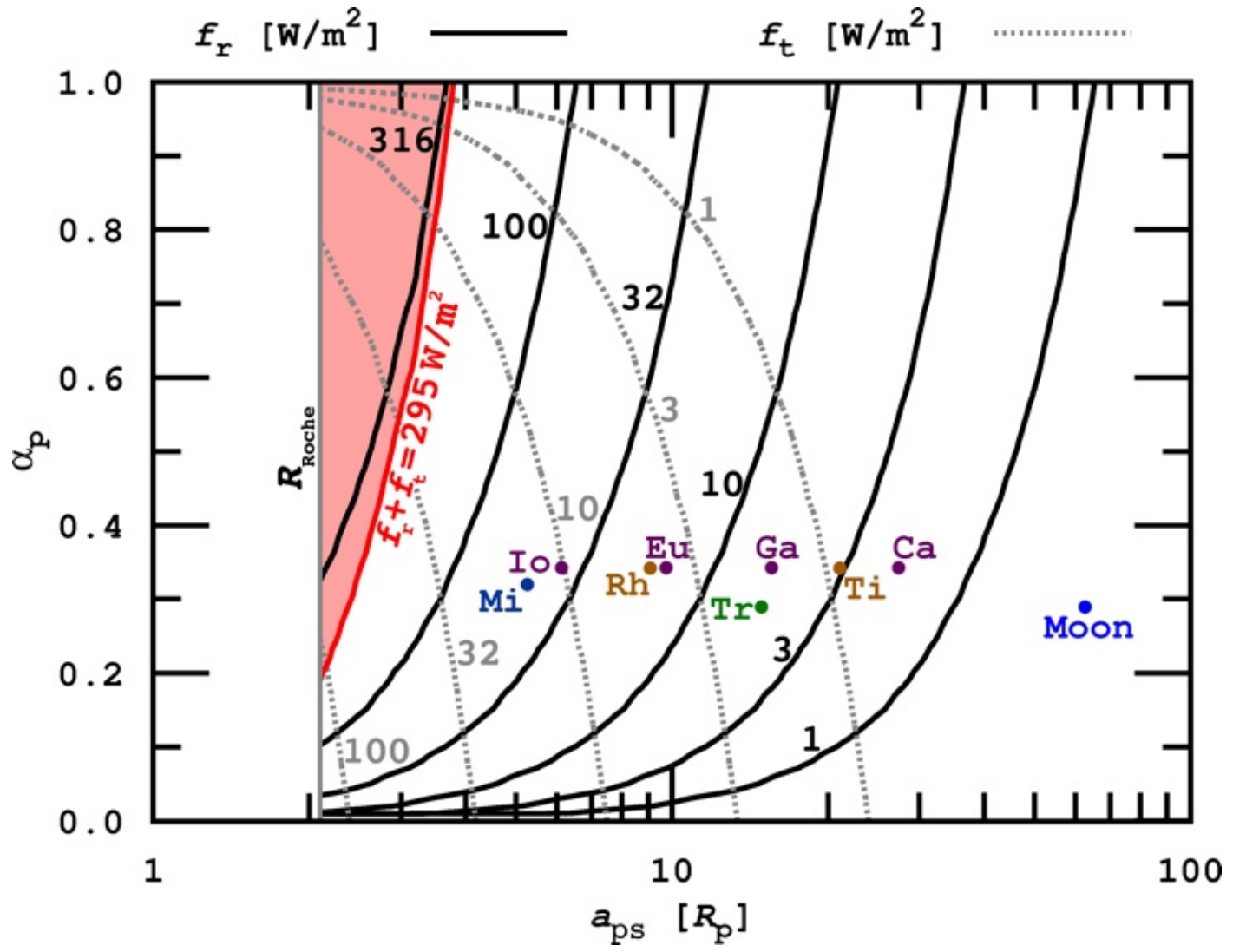

**FIG. 2.** Contours of constant planetary flux on an exomoon as a function of the planet-satellite semi-major axis $a_{ps}$ and the planet's bond albedo $\alpha_p$. The planet-moon binary orbits at 1AU from a Sun-like host star. Values depict the maximum possible irradiation in terms of orbital alignment, i.e., on the subplanetary point on the moon, and when the moon is over the substellar point of the planet. For $\alpha_p \approx 0.1$ contours of equal $f_r$ and $f_t$ intersect, i.e., both contributions are equal. An additional contour is added at 295W/m$^2$, where the sum of $f_r$ and $f_t$ induces a runaway greenhouse on an Earth-sized moon. Some examples from the Solar System are given: Miranda (Mi), Io, Rhea (Rh), Europa (Eu), Triton (Tr), Ganymede (Ga), Titan (Ti), Callisto (Ca), and Earth's moon (Moon).

### 3.1.3 The circumstellar habitable zone of exomoons

We next transform the combined stellar and planetary flux into a correction for the IHZ, for which the boundaries are proportional to $L_*^{1/2}$ (Selsis et al. 2007). This correction is easily derived if we restrict the problem to just the direct and the reflected starlight. Then, we can define an "effective luminosity" $L_{eff}$ that is the sum of the direct starlight plus the orbit-averaged reflected light. We ignore the thermal contribution as its spectral energy distribution will be much different from the star and, as shown below, the thermal component is the smallest for most cases. Our IHZ corrections are therefore only lower limits. From Eqs. (3), (13), and (16) one can show that

$$L_{eff} = L_* \Big( 1 + \frac{\alpha_p R_p^2}{8 a_{ps}^2} \Big) , \tag{17}$$

where we have averaged over the moon's orbital period. For realistic moon orbits, this correction amounts to 1% at most for high $\alpha_p$ and small $a_{ps}$. For planets orbiting F dwarfs near the outer edge of the IHZ, a moon could be habitable about 0.05AU farther out due to the reflected planetary light. In Fig. 3, we show the correction factor for the inner and outer boundaries of the IHZ due to reflected light as a function of $\alpha_p$ and $a_{ps}$.

### 3.1.4 Combined stellar and planetary illumination

With Eqs. (3), (12), and (16), we have derived the stellar and planetary contributions to the irradiation of a tidally locked moon in an inclined, circular orbit around the planet, where the orbit of the planet-moon duet around the star is eccentric. Now, we consider a satellite's total illumination

$$f_s(t) = f_*(t) + f_t(t) + f_r(t) \ . \tag{18}$$

For an illustration of Eq. (18), we choose a moon that orbits its Jupiter-sized host planet at the same distance as Europa orbits Jupiter. The planet-moon duet is in a 1AU orbit around a Sun-like star, and we arbitrarily choose a temperature difference of d$T$ = 100K between the two planetary hemispheres. Equation (18) does not depend on $M_s$ or $R_s$, so our irradiation model is not restricted to either the Earth-sized or the Super-Ganymede prototype moon.

In Fig. 4 we show $f_s(t)$ as well as the stellar and planetary contributions for four different locations on the moon's surface. For all panels, the planet-moon duet is at the beginning of its revolution around the star, i.e. $\mathfrak{M}_{*p} = 0$, and we set $i = 0$. Although $\mathfrak{M}_{*p}$ would slightly increase during one orbit of the moon around the planet, we fix it to zero, so the moon starts and finishes over the illuminated hemisphere of the planet (similar to Fig. 1). The upper left panel depicts the subplanetary point, with a pronounced eclipse around $\varphi_{ps} = 0.5$. At a position 45° counterclockwise along the equator (upper right panel), the stellar contribution is shifted in phase, and $f_t$ as well as $f_r$ are diminished in magnitude (note the logarithmic scale!) by a factor cos(45°). In the lower row, where $\phi$ = 90°, $\theta$ = 0° (lower left panel) and $\phi$ = 180°, $\theta$ = 80° (lower right panel), there are no planetary contributions. The eclipse trough has also disappeared because the star's occultation by the planet cannot be seen from the antiplanetary hemisphere.

For Fig. 5, we assume a similar system, but now the planet-moon binary is at an orbital phase $\varphi_{*p} = 0.5$, corresponding to $\mathfrak{M}_{*p} = \pi$, around the star. We introduce an eccentricity $e_{*p} = 0.3$ as well as an inclination of 45° between the two orbital planes. The first aspect shifts the stellar and planetary contributions by half an orbital phase with respect to Fig. 4. Considering the top view of the system in Fig. 1, this means eclipses should now occur when the moon is to the left of the planet because the star is to the right, "left" here meaning $\varphi_{ps} = 0$. However, the non-zero inclination lifts the moon out of the planet's shadow (at least for this particular orbital phase around the star), which is why the eclipse trough disappears. Due to the eccentricity, stellar irradiation is now lower because the planet-moon binary is at apastron. An illustration of the corresponding star-planet orbital configuration is shown in the pole view (left panel) of Fig. 6.

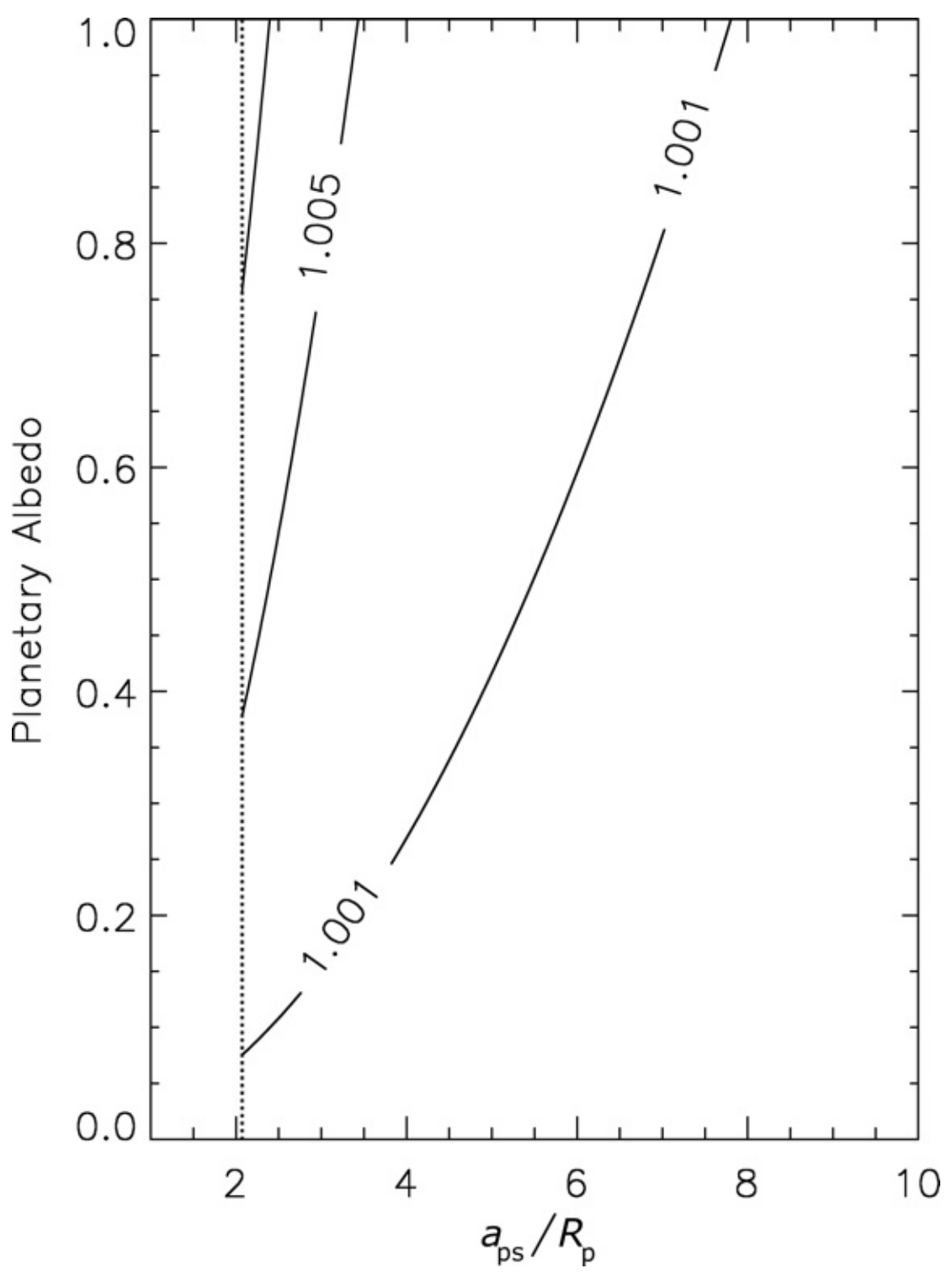

**FIG. 3.** Contours of the correction factor for the limits of the IHZ for exomoons, induced by the star's reflected light from the planet. Since we neglect the thermal component, values are lower limits. The left-most contour signifies 1.01. The dotted vertical line denotes the Roche lobe.

The eccentricity-driven cooling of the moon is enhanced on its northern hemisphere, where the inclination induces a winter. Besides, our assumption that the moon is in the planet's equatorial plane is equivalent to $i = \psi_p$; thus the planet also experiences northern winter. The lower right panel of Fig. 5, where $\phi = 180°$ and $\theta = 80°$, demonstrates a novel phenomenon, which we call an "antiplanetary winter on the moon". On the satellite's antiplanetary side there is no illumination from the planet (as in the lower two panels of Fig. 4); and being close enough to the pole, at $\theta > 90° - i$ for this occasion of northern winter, there will be no irradiation from the star either, during the whole orbit of the moon around the planet. In Fig. 6, we depict this constellation in the edge view (right panel). Note that antiplanetary locations close to the moon's northern pole receive no irradiation at all, as indicated by an example at $\phi = 180°$, $\theta = 80°$ (see arrow). Of course there is also a "proplanetary winter" on the moon, which takes place just at the same epoch but on the proplanetary hemisphere on the moon. The opposite effects are the "proplanetary summer", which occurs on the proplanetary side of the moon at $\mathfrak{M}_{*p} = 0$, at least for this specific configuration in Fig. 5, and the "antiplanetary summer".

Finally, we compute the average surface flux on the moon during one stellar orbit. Therefore, we first integrate $d\varphi_{ps} f_s(\varphi_{ps})$ over $0 \leq \varphi_{ps} \leq 1$ at an initial phase in the planet's orbit about the star ($\varphi_{*p} = 0$), which yields the area under the solid lines in Fig. 4. We then step through $\approx 50$ values for $\varphi_{*p}$ and again integrate the total flux. Finally, we average the flux over one orbit of the planet around the star, which gives the orbit-averaged flux $F_s(\phi,\theta)$ on the moon. In Fig. 7, we plot these values as surface maps of a moon in four scenarios. The two narrow panels to the right of each of the four major panels show the averaged flux for $-1/4 \leq \varphi_{*p} \leq +1/4$ and $+1/4 \leq \varphi_{*p} \leq +3/4$, corresponding to northern summer (ns) and southern summer (ss) on the moon, respectively.

In the upper left panel, the two orbits are coplanar. Interestingly, the subplanetary point at $\phi = 0 = \theta$ is the "coldest" spot along the equator (if we convert the flux into a temperature) because the moon passes into the shadow of the planet when the star would be at zenith over the subplanetary point. Thus, the stellar irradiation maximum is reduced (see the upper left panel in Fig. 4). The contrast between polar and equatorial irradiation, reaching from 0 to $\approx$440W/m$^2$, is strongest in this panel. In the upper right panel, the subplanetary point has turned into the "warmest" location along the equator. On the one hand, this is due to the inclination of 22.5°, which is why the moon does not transit behind the planet for most of the orbital phase around the star. On the other hand, this location gets slightly more irradiation from the planet than any other place on the moon. In the lower left panel, the average flux contrast between equatorial and polar illumination has decreased further. Again, the subplanetary point is slightly warmer than the rest of the surface. In the lower right panel, finally, where the moon's orbital inclination is set to 90°, the equator has become the coldest region of the moon, with the subplanetary point still being the warmest location along 0° latitude.

While the major panels show that the orbit-averaged flux contrast decreases with increasing inclination, the side panels indicate an increasing irradiation contrast between seasons. Exomoons around host planets with obliquities similar to that of Jupiter with respect to the Sun ($\psi_p \approx 0$) are subject to an irradiation pattern corresponding to the upper left panel of Fig. 7. The upper right panel depicts an irradiation pattern of exomoons around planets with obliquities similar to Saturn (26.7°) and is qualitatively in good agreement with the yearly illumination pattern of Titan as simulated by Mitchell (2012, see his Fig. 1c). Exomoons around a planet with a Uranus-like obliquity (97.9°) will have an irradiation similar to the lower right panel.

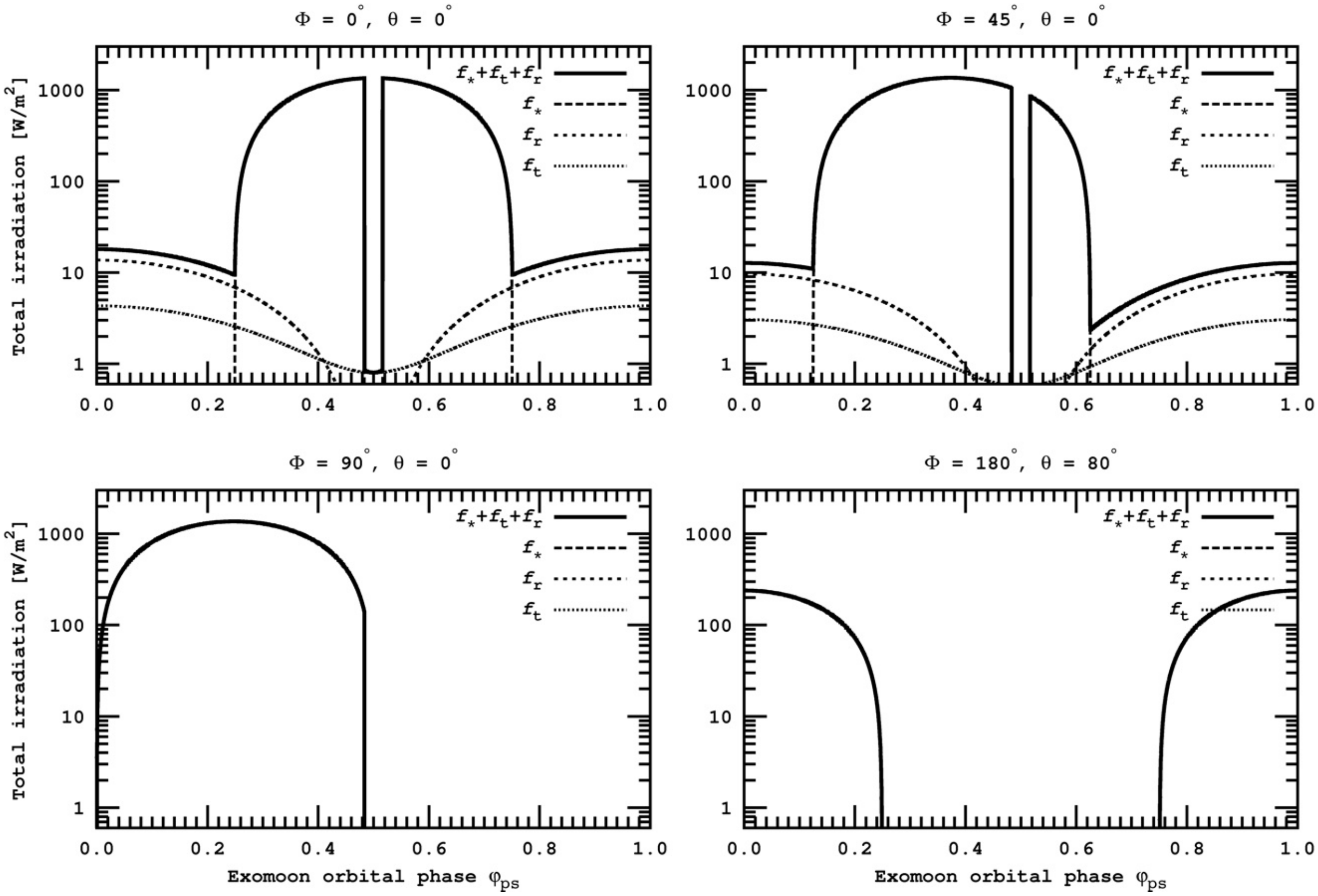

**FIG. 4.** Stellar and planetary contributions to the illumination of our prototype moon as a function of orbital phase $\varphi_{ps}$. Tiny dots label the thermal flux from the planet ($f_t$), normal dots the reflected stellar light from the planet ($f_r$), dashes the stellar light ($f_*$), and the solid line is their sum. The panels depict different longitudes and latitudes on the moon's surface. The upper left panel is for the subplanetary point, the upper right 45° counterclockwise along the equator, the lower left panel shows a position 90° counterclockwise from the subplanetary point, and the lower right is the antiplanetary point.

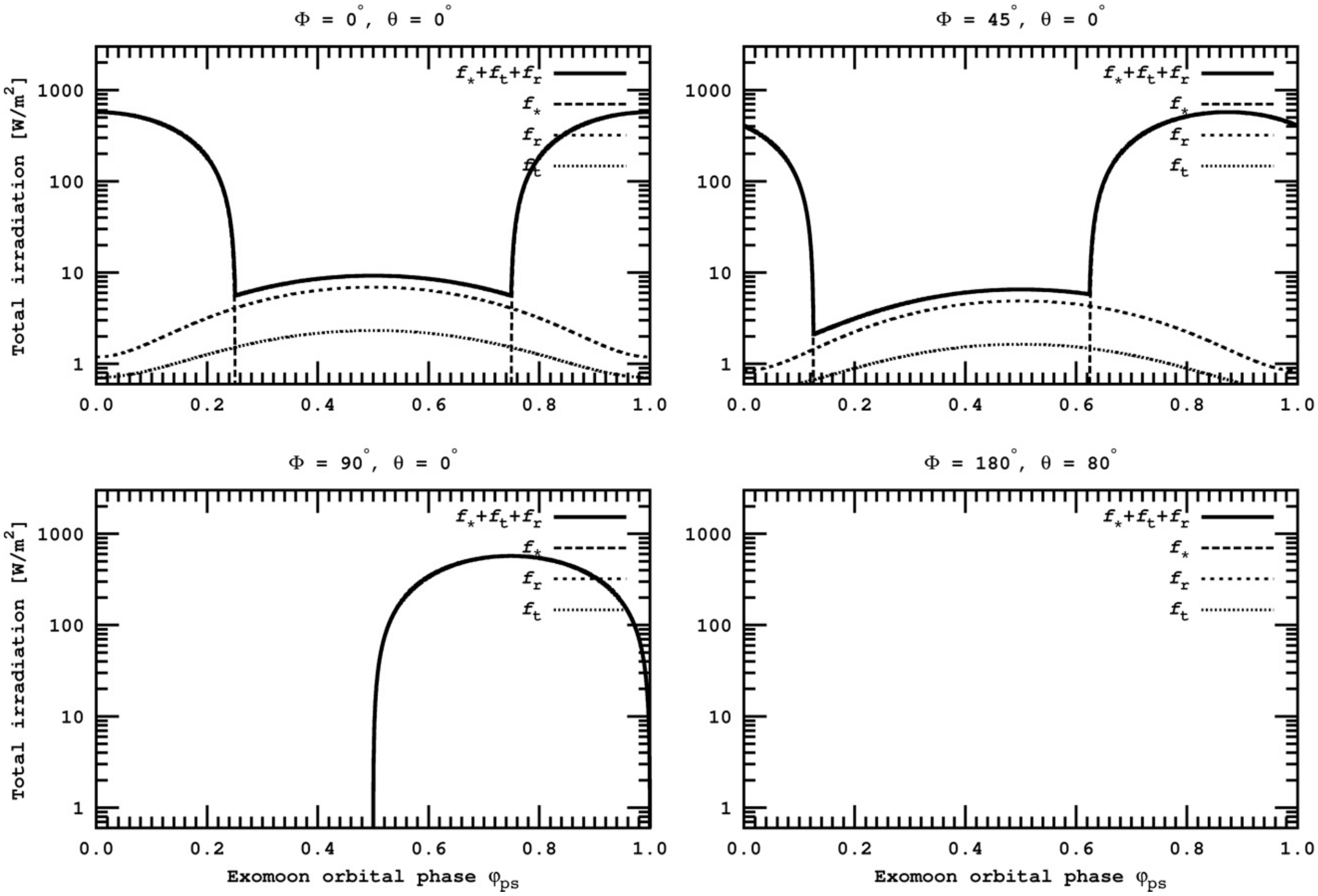

**FIG. 5.** Stellar and planetary contributions to the illumination of our prototype moon as in Fig. 4 but at a stellar orbital phase $\varphi_{*p} = 0.5$ in an eccentric orbit ($e_{*p} = 0.3$) and with an inclination $i = \pi/4 \mathrel{\hat{=}} 45°$ of the moon's orbit against the circumstellar orbit.

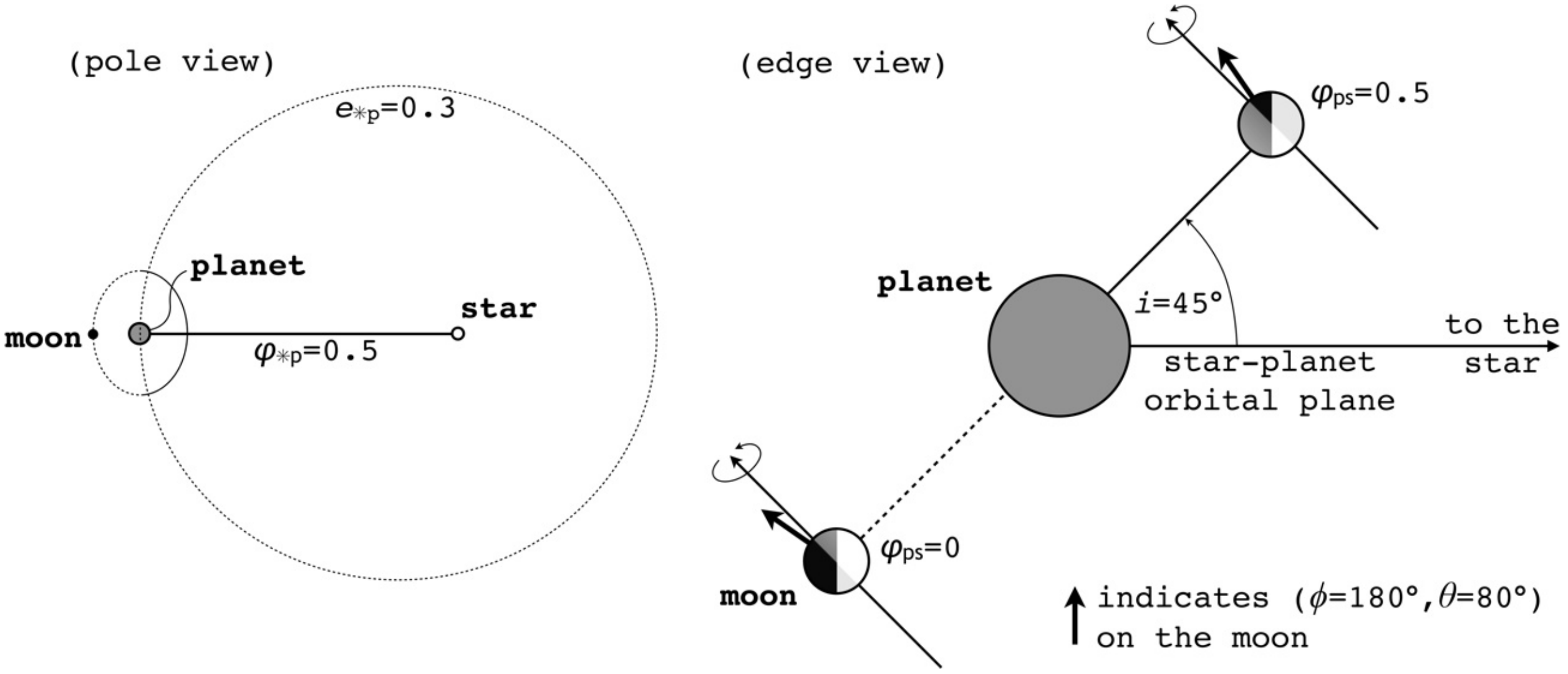

**FIG. 6.** Illustration of the antiplanetary winter on the moon with the same orbital elements as in Fig. 5. The arrow in the edge view panel indicates the surface normal at $\phi$=180°, $\theta$=80°, i.e., close to pole and on the antiplanetary side of the moon. For all orbital constellations of the moon around the planet ($\varphi_{ps}$ going from 0 to 1), this location on the moon receives neither irradiation from the star nor from the planet (see lower right panel in Fig. 5). Shadings correspond to the same irradiation patterns as in Fig. 1.

The typical orbit-averaged flux between 300 and 400W/m$^2$ in Fig. 7 is about a quarter of the solar constant. This is equivalent to an energy redistribution factor of 4 over the moon's surface (Selsis et al. 2007), indicating that climates on exomoons with orbital periods of a few days (in this case 3.55d, corresponding to Europa's orbit about Jupiter) may be more similar to those of freely rotating planets rather than to those of planets that are tidally locked to their host star.

### 3.2 Tidal heating

Tidal heating is an additional source of energy on moons. Various approaches for the description of tidal processes have been established. Two of the most prominent tidal theories are the "constant-time-lag" (CTL) and the "constant-phase-lag" (CPL) models. Their merits and perils have been treated extensively in the literature (Ferraz-Mello et al. 2008; Greenberg 2009; Efroimsky & Williams 2009; Hansen 2010; Heller et al. 2010,2011b; Lai 2012) and it turns out that they agree for low eccentricities. To begin with, we arbitrarily choose the CTL model developed by Hut (1981) and Leconte et al. (2010) for the computation of the moon's instantaneous tidal heating, but we will compare predictions of both CPL and CTL theory below. We consider a tidal time lag $\tau_s$ = 638s, similar to that of Earth (Lambeck 1977; Neron de Surgy & Laskar 1997), and an appropriate second-order potential Love number of $k_{2,s}$ = 0.3 (Henning et al. 2009).

In our two-body system of the planet and the moon, tidal heating on the satellite, which is assumed to be in equilibrium rotation and to have zero obliquity against the orbit around the planet, is given by

$$\dot{E}_{\rm tid,s}^{\rm eq} = \frac{Z_{\rm s}}{\beta^{15}(e_{\rm ps})} \left[ f_1(e_{\rm ps}) - \frac{f_2^2(e_{\rm ps})}{f_5(e_{\rm ps})} \right] \ , \qquad (19)$$

where

$$Z_{\rm s} \equiv 3G^2 k_{2,{\rm s}} M_{\rm p}^2 (M_{\rm p} + M_{\rm s}) \frac{R_{\rm s}^5}{a_{\rm ps}^9} \ \tau_{\rm s} \qquad (20) \qquad \text{and} \qquad \beta(e_{\rm ps}) = \sqrt{1 - e_{\rm ps}^2} \ ,$$

$$\begin{aligned} f_1(e_{\rm ps}) &= 1 + \frac{31}{2} e_{\rm ps}^2 + \frac{255}{8} e_{\rm ps}^4 + \frac{185}{16} e_{\rm ps}^6 + \frac{25}{64} e_{\rm ps}^8 \ , \\ f_2(e_{\rm ps}) &= 1 + \frac{15}{2} e_{\rm ps}^2 + \frac{45}{8} e_{\rm ps}^4 + \frac{5}{16} e_{\rm ps}^6 \ , \\ f_5(e_{\rm ps}) &= 1 + 3 e_{\rm ps}^2 + \frac{3}{8} e_{\rm ps}^4 \qquad (21). \end{aligned}$$

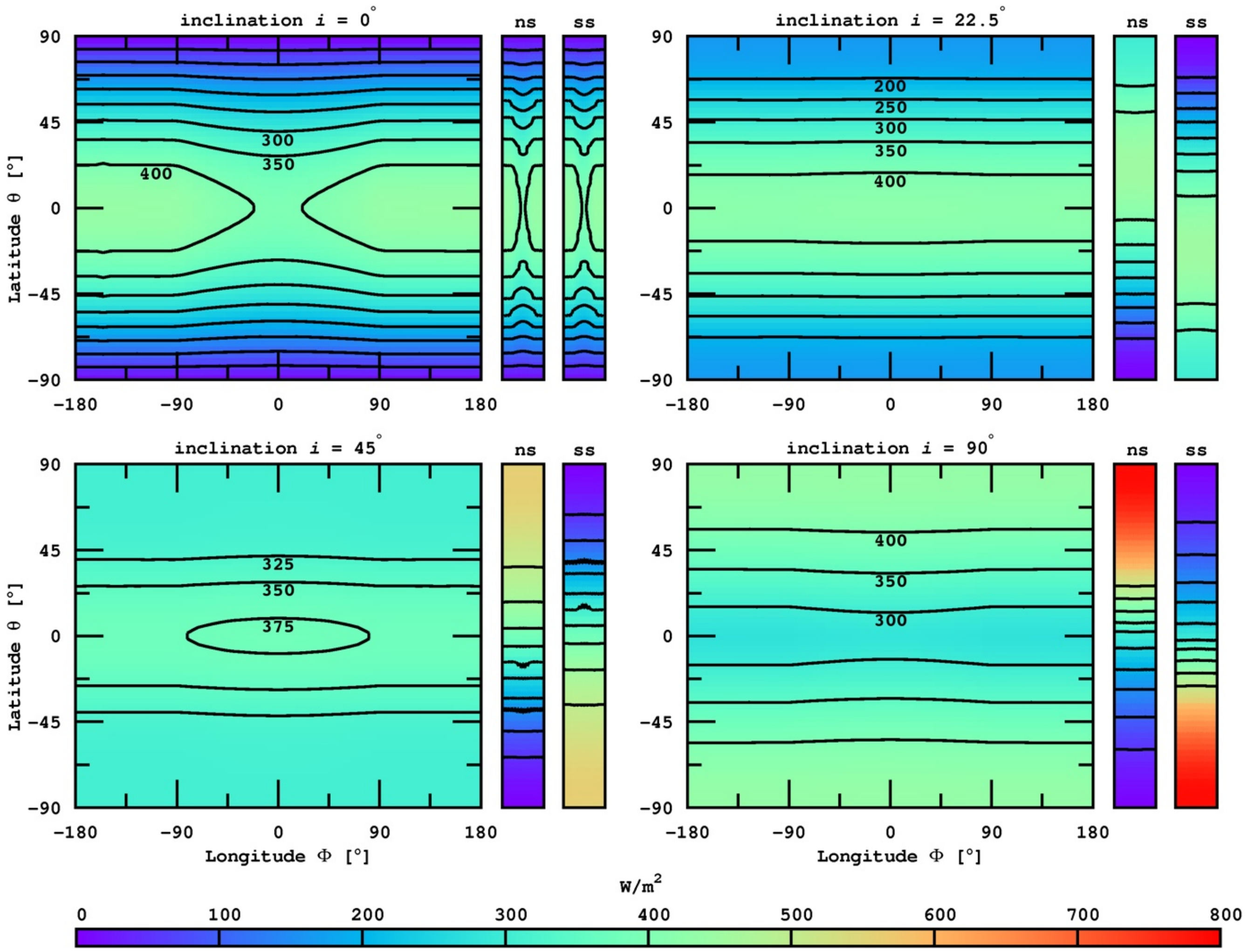

**FIG. 7.** Illumination of our prototype exomoon (in W/m$^2$) averaged over the orbit of the planet-moon duet around their host star. Major panels present four different orbital inclinations: $i = 0°$ (upper left), $i = 22.5°$ (upper right), $i = 45°$ (lower left), and $i = 90°$ (lower right). The two bars beside each major panel indicate averaged flux for the northern summer (ns) and southern summer (ss) on the moon. Contours of constant irradiation are symmetric about the equator; some values are given.

Here, $G$ is Newton's gravitational constant, and $M_{\rm p}$ is the planet's mass. The contribution of tidal heating to the moon's energy flux can be compared to the incoming irradiation when we divide $\dot{E}_{\rm tid,s}^{\rm eq}$ by the surface of the moon and define its surface tidal heating $h_{\rm s} \equiv \dot{E}_{\rm tid,s}^{\rm eq}/(4\pi R_{\rm s}^2)$. We assume that $h_{\rm s}$ is emitted uniformly through the satellite's surface.

To stress the importance of tidal heating on exomoons, we show the sum of $h_{\rm s}$ and the absorbed stellar flux for four star-planet-moon constellations as a function of $e_{\rm ps}$ and $a_{\rm ps}$ in Fig. 8. In the upper row, we consider our Earth-like prototype moon, in the lower row the Super-Ganymede. The left column corresponds to a Jupiter-like host planet, the right column to a Neptune-like planet, both at 1AU from a Sun-like star. Contours indicate regions of constant energy flux as a function of $a_{\rm ps}$ and $e_{\rm ps}$. Far from the planet, illustrated by a white area at the right in each plot, tidal heating is negligible, and the total heat flux on the moon corresponds to an absorbed stellar flux of 239W/m$^2$, which is equal to Earth's absorbed flux. The right-most contour in each panel depicts a contribution by tidal heating of 2W/m$^2$, which corresponds to Io's tidal heat flux (Spencer et al. 2000). Satellites left of this line will not necessarily experience enhanced volcanic activity, since most of the dissipated energy would go into water oceans of our prototype moons, rather than into the crust as on Io. The blue contour corresponds to a tidal heating of 10W/m$^2$, and the red line demarcates the transition into a runaway greenhouse, which occurs at 295W/m$^2$ for the Earth-mass satellite and at 266W/m$^2$ for the Super-Ganymede. For comparison, we show the positions of some prominent moons in the Solar System, where $a_{\rm ps}$ is measured in radii of the host planet. Intriguingly, both the Earth-like exomoon and the Super-Ganymede, orbiting either a Jupiter- or a Neptune-mass planet, would be habitable in a Europa- or Miranda-like orbit (in terms of fractional planetary radius and eccentricity), while they would enter a runaway greenhouse state in an Io-like orbit.

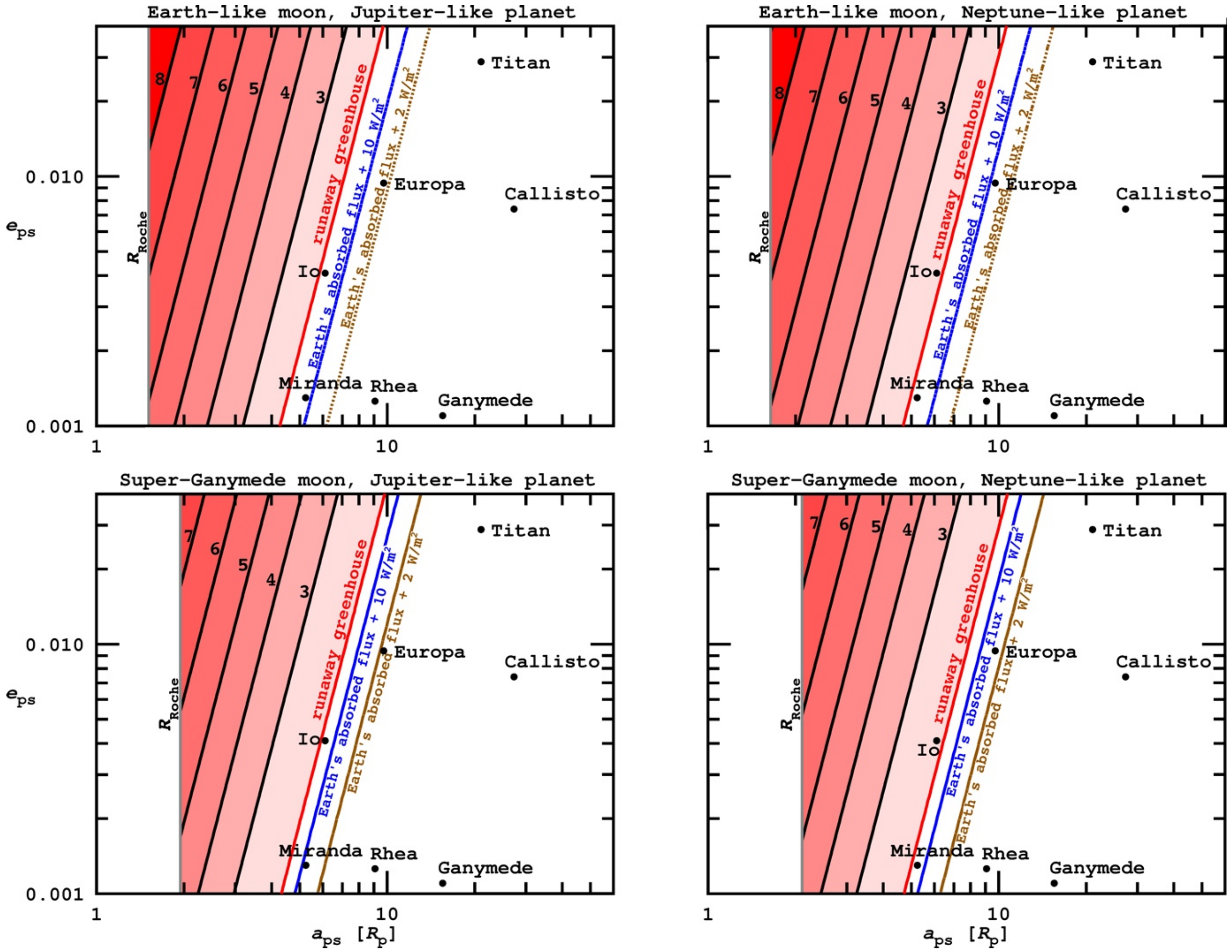

**FIG. 8.** Contours of summed absorbed stellar irradiation and tidal heating (in logarithmic units of W/m$^2$) as a function of semi-major axis $a_{ps}$ and eccentricity $e_{ps}$ on an Earth-like (upper row) and a Super-Ganymede (lower row) exomoon. In the left panels, the satellite orbits a Jupiter-like planet, in the right panels a Neptune-mass planet, in both cases at 1AU from a Sun-like host star. In the white area at the right, tidal heating is negligible and absorbed stellar flux is 239W/m$^2$. The right-most contours in each panel indicate Io's tidal heat flux of 2W/m$^2$, a tidal heating of 10W/m$^2$, and the critical flux for the runaway greenhouse (295W/m$^2$ for the Earth-like moon and 266W/m$^2$ for the Super-Ganymede). Positions of some massive satellites in the Solar System are shown for comparison.

The Roche radii are $\approx 1.5 R_p$ for an Earth-type moon about a Jupiter-like planet (upper left panel in Fig. 8), $\approx 1.6 R_p$ for an Earth-type moon about a Neptune-mass planet (upper right), $\approx 1.9 R_p$ for the Super-Ganymede about a Jupiter-class host (lower left), and $\approx 2.1 R_p$ for the Super-Ganymede orbiting a Neptune-like planet (lower right). The extreme tidal heating rates in the red areas may not be realistic because we assume a constant time lag $\tau_s$ of the satellite's tidal bulge and ignore its dependence on the driving frequency as well as its variation due to the geological processes that should appear at such enormous heat fluxes.

In Fig. 8, irradiation from the planet is neglected; thus decreasing distance between planet and moon goes along with increasing tidal heating only. An Earth-like exomoon (upper panels) could orbit as close as $\approx 5 R_p$, and tidal heating would not induce a runaway greenhouse if $e_{ps} \lesssim 0.001$. If its orbit has an eccentricity similar to Titan ($e_{ps} \approx 0.03$), then the orbital separation needed to be $\gtrsim 10 R_p$ to prevent a runaway greenhouse. Comparison of the left and right panels shows that for a more massive host planet (left) satellites can be slightly closer and still be habitable. This is because we draw contours over the fractional orbital separation $a_{ps}/R_p$ on the abscissa. In the left panels $10 R_p \approx 7 \times 10^5$km, whereas in the right panel $10 R_p \approx 2.5 \times 10^5$km. When plotting over $a_{ps}/R_p$, this discrepancy is somewhat balanced by the much higher mass of the host planet in the left panel and the strong dependence of tidal heating on $M_p$ (see Eq. 20).

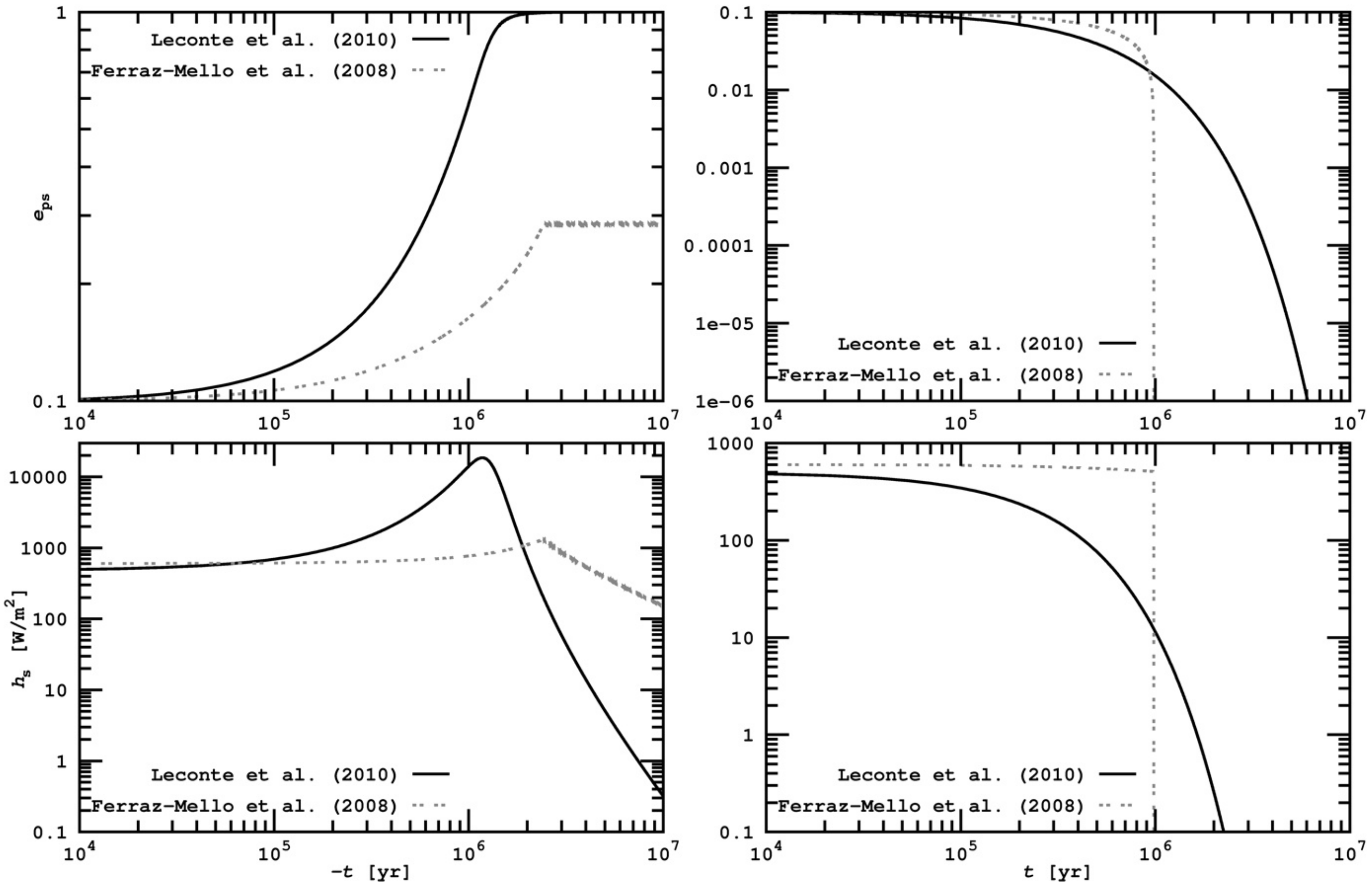

**FIG. 9.** Evolution of the orbital eccentricity (upper row) and the moon's tidal heating (lower row) following the two-body tidal models of Leconte et al. (2010) (a "constant-time-lag" model, solid line) and Ferraz-Mello et al. (2008) (a "constant-phase-lag" model, dashed line). Initially, an Earth-sized moon is set in an eccentric orbit ($e_{ps} = 0.1$) around a Jupiter-mass planet at the distance in which Europa orbits Jupiter. In the left panels evolution is backward, in the right panels into the future. Both tidal models predict that free eccentricities are eroded and tidal heating ceases after <10Myr.

### 3.2.1 Planet-moon orbital eccentricity

Tidal heating on the moon will only be significant, if $e_{ps} \neq 0$. If the eccentricity is *free*, that is, the moon is not significantly perturbed by other bodies, then $e_{ps}$ will change with time due to tidal damping, and orbit-averaged equations can be applied to simulate the tidal evolution. For those cases considered here, $de_{ps}/dt < 0$. Similar to the approach used by Heller et al. (2011b), we apply a CTL model (Leconte et al. 2010) and a CPL model (Ferraz-Mello et al. 2008) to compute the tidal evolution, albeit here with a focus on exomoons rather than on exoplanets.

In Fig. 9, we show the evolution of our Earth-sized prototype moon in orbit around a Jupiter-like planet. The upper two panels show the change of eccentricity $e_{ps}(t)$, the lower two panels of tidal surface heating $h_s(t)$. In the left panels, evolution is backward until $-10^7$yr; in the right panels, evolution is forwards until $+10^7$yr. The initial eccentricity is set to 0.1, and for the distance between the planet and the moon we choose the semi-major axis of Europa around Jupiter. Going backward, eccentricity increases[8] as does tidal heating. When $e_{ps} \rightarrow 1$ at $\approx -1$Myr, the moon is almost moving on a line around the planet, which would lead to a collision (or rather to an ejection because we go backward in time). At this time, tidal heating has increased from initially 600W/m$^2$ to 1300W/m$^2$. The right panels show that free eccentricities will be damped to zero within <10Myr and that tidal heating becomes negligible after ≈2Myr.

Although $e_{ps}$ is eroded in <10Myr in these two-body simulations, eccentricities can persists much longer, as the cases of Io around Jupiter and Titan around Saturn show. Their eccentricities are not free, but they are *forced*, because they are excited by interaction with other bodies. The origin of Titan's eccentricity $e_{ps} = 0.0288$ is still subject to debate. As shown by Sohl et al. (1995), tidal dissipation would damp it on timescales shorter than the age of the Solar System. It could only be primordial if the moon had a methane or hydrocarbon ocean that is deeper than a few kilometers. However, surface observations by the *Cassini Huygens* lander negated this assumption. Various other possibilities have been discussed, such as collisions or close encounters with massive bodies (Smith et al. 1982; Farinella et al. 1990) and a capture of Titan by Saturn ≪4.5Gyr ago. The origin of Io's eccentricity $e_{ps} = 0.0041$ lies in the moon's resonances with Ganymede and Callisto (Yoder 1979). Such resonances may also appear among exomoons.

[8] For the CPL model, $e_{ps}$ converges to 0.285 for $-t \gtrsim 2$Myr. This result is not physical but owed to the discontinuities induced by the phase lags, in particular by $\varepsilon_{1,s}$ and $\varepsilon_{1,p}$ (for details see Ferraz-Mello et al. 2008; Heller et al. 2011b). We have tried other initial eccentricities, which all led to this convergence.

Using the publicly available *N*-body code *Mercury* (Chambers 1999)[9], we performed *N*-body experiments of a hypothetical satellite system to find out whether forced eccentricities can drive a tidal greenhouse in a manner analogous to the volcanic activity of Io. We chose a Jupiter-mass planet with an Earth-mass exomoon at the same distance as Europa orbits Jupiter, placed a second exomoon at the 2:1 external resonance, and integrated the resulting orbital evolution. In one case, the second moon had a mass equal to that of Earth, in the other case a mass equal to that of Mars. For the former case, we find the inner satellite could have its eccentricity pumped to 0.09 with a typical value of 0.05. For the latter, the maximum eccentricity is 0.05 with a typical value near 0.03. Although these studies are preliminary, they suggest that massive exomoons in multiple configurations could trigger a runaway greenhouse, especially if the moons are of Earth-mass, and that their circumstellar IHZ could lie further away from the star. A comprehensive study of configurations that should also include Cassini states and damping to the fixed-point solution is beyond the scope of this study but could provide insight into the likelihood that exomoons are susceptible to a tidal greenhouse.

## 4. Orbits of habitable exomoons

By analogy with the circumstellar habitable zone for planets, we can imagine a minimum orbital separation between a planet and a moon to let the satellite be habitable. The range of orbits for habitable moons has no outer edge, except that Hill stability must be ensured. Consequently, habitability of moons is only constrained by the inner edge of a circumplanetary habitable zone, which we call the "habitable edge". Moons inside the habitable edge are in danger of running into a greenhouse by stellar and planetary illumination and/or tidal heating. Satellites outside the habitable edge with their host planet in the circumstellar IHZ are habitable by definition.

Combining the limit for the runaway greenhouse from Section 2.2 with our model for the energy flux budget of extrasolar moons from Section 3, we compute the orbit-averaged global flux $\bar{F}_{\rm s}^{\rm glob}$ received by a satellite, which is the sum of the averaged stellar $(\bar{f}_*)$, reflected $(\bar{f}_{\rm r})$, thermal $(\bar{f}_{\rm t})$, and tidal heat flux ($h_{\rm s}$). Thus, in order for the moon to be habitable

$$\begin{aligned} F_{\rm RG} > \bar{F}_{\rm s}^{\rm glob} &= \bar{f}_* + \bar{f}_{\rm r} + \bar{f}_{\rm t} + h_{\rm s} \\ &= \frac{L_*(1-\alpha_{\rm s})}{16\pi a_{*{\rm p}}^2\sqrt{1-e_{*{\rm p}}^2}}\left(1+\frac{\pi R_{\rm p}^2\alpha_{\rm p}}{2a_{\rm ps}^2}\right) + \frac{R_{\rm p}^2\sigma_{\rm SB}(T_{\rm p}^{\rm eq})^4}{a_{\rm ps}^2}\frac{(1-\alpha_{\rm s})}{4} + h_{\rm s} \quad , \end{aligned} \tag{22}$$

where the critical flux for a runaway greenhouse $F_{\rm RG}$ is given by Eq. (1), the tidal heating rate $h_{\rm s} \equiv \dot{E}_{\rm tid,s}^{\rm eq}/(4\pi R_{\rm s}^2)$ by Eq. (19), and the planet's thermal equilibrium temperature $T_{\rm p}^{\rm eq}$ can be determined with Eq. (11) and using $\mathrm{d}T = 0$. Note that the addend "1" in brackets implies that we do not consider reduction of stellar illumination due to eclipses. This effect is treated in a companion paper (Heller 2012).

In Fig. 10, we show the $F_{\rm RG} = \bar{F}_{\rm s}^{\rm glob}$ orbits of both the Earth-like (blue lines) and the Super-Ganymede (black lines) prototype moon as a function of the planet-moon semi-major axis and the mass of the giant host planet, which orbits the Sun-like star at a distance of 1AU.[10] Contours are plotted for various values of the orbital eccentricity, which means that orbits to the left of a line induce a runaway greenhouse for the respective eccentricity of the actual moon. These innermost, limiting orbits constitute the circumplanetary habitable edges.

When the moon is virtually shifted toward the planet, then illumination from the planet and tidal heating increase, reaching $F_{\rm RG}$ at some point. With increasing eccentricity, tidal heating also increases; thus $F_{\rm RG}$ will be reached farther away from the planet. Blue lines appear closer to the planet than black contours for the same eccentricity, showing that more massive moons can orbit more closely to the planet and be prevented from becoming a runaway greenhouse. This is a purely atmospheric effect, determined only by the moon's surface gravity $g_s(M_s, R_s)$ in Eq. (1). We also see that, for a fixed eccentricity, moons can orbit closer to the planet – both in terms of fractional and absolute units – if the planet's mass is smaller.

[9] Download via www.arm.ac.uk/~jec .

[10] For consistency, we computed the planetary radius (used to scale the abscissa) as a function of the planet's mass (ordinate) by fitting a high-order polynomial to the Fortney et al. (2007) models for a giant planet at 1AU from a Sun-like star (see line 17 in their Table 4).

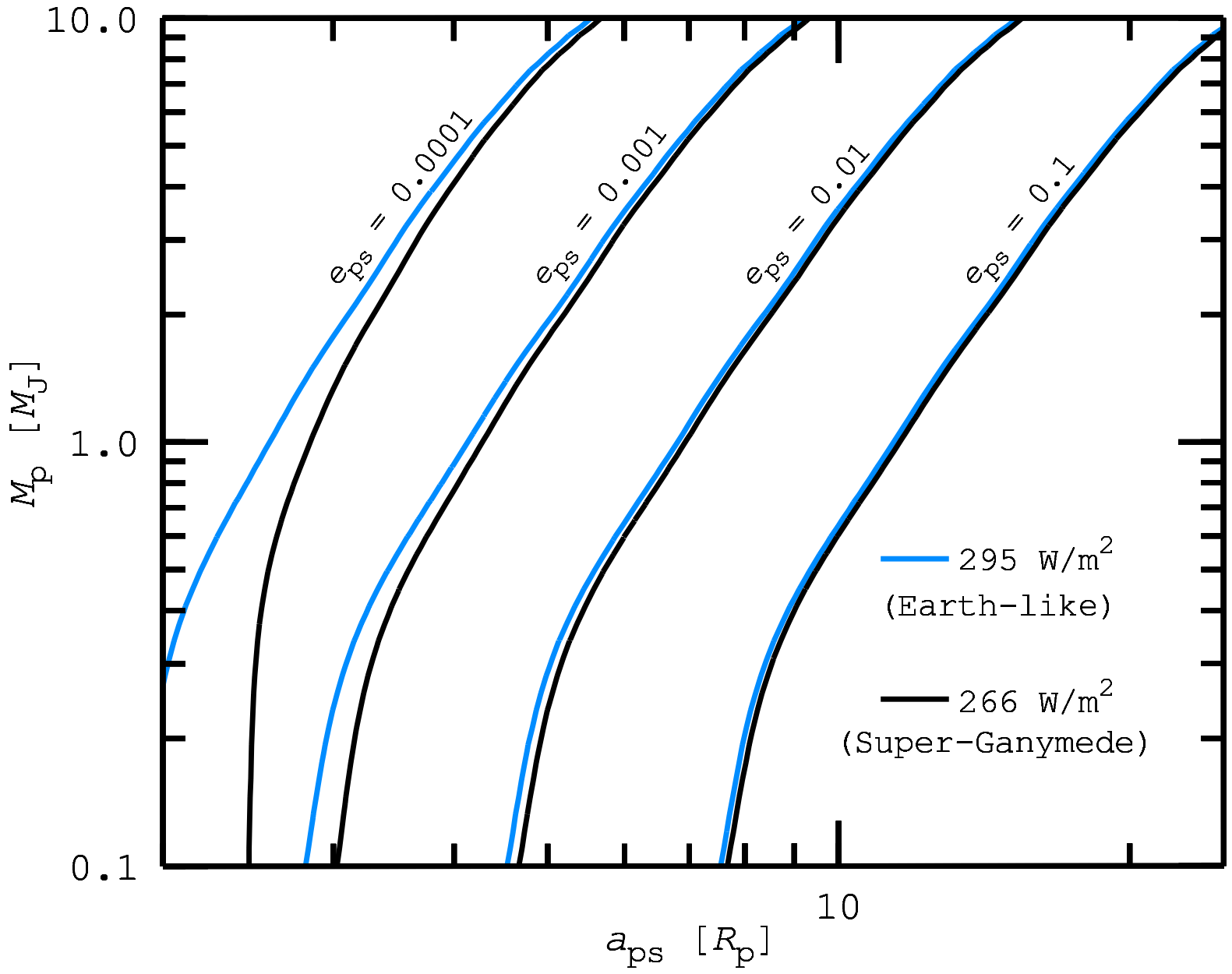

**FIG. 10.** Innermost orbits to prevent a runaway greenhouse, i.e., the "habitable edges" of an Earth-like (blue lines) and a Super-Ganymede (black lines) exomoon. Their host planet is at 1AU from a Sun-like star. Flux contours for four eccentricities of the moons' orbits from $e_{ps} = 10^{-4}$ to $e_{ps} = 10^{-1}$ are indicated. The larger $e_{ps}$, the stronger tidal heating and the more distant from the planet will the critical flux be reached.

Estimating an exomoon's habitability with this model requires a well-parametrized system. With the current state of technology, stellar luminosity $L_*$ and mass $M_*$ can be estimated by spectral analysis and by using stellar evolution models. In combination with the planet's orbital period $P_{*p}$, this yields $a_{*p}$ by means of Kepler's third law. The planetary mass $M_p$ could be measured with the radial-velocity (RV) method and assuming it is sufficiently larger than the moon's mass. Alternatively, photometry could determine the mass ratios $M_p/M_*$ and $M_s/M_p$. Thus, if $M_s$ were known from spectroscopy and stellar evolution models, then all masses were accessible (Kipping 2010). Combining RV and photometry, the star-planet orbital eccentricity $e_{*p}$ can be deduced (Mislis et al. 2012); and by combination of TTV and TDV, it is possible to determine $M_s$ as well as $a_{ps}$ (Kipping 2009a). Just like the planetary radius, the moon's radius $R_s$ can be determined from photometric transit observations if its transit can directly be observed. The satellite's second-order tidal Love number $k_{2,s}$ and its tidal time lag $\tau_s$, however, would have to be assumed. For this purpose, Earth or Solar System moons could serve as reference bodies. If the age of the system (i.e., of the star) was known or if the evolution of the satellite's orbit due to tides could be observed, then tidal theory could give a constraint on the product of $k_{2,s}$ and $\tau_s$ (or $k_{2,s}/Q_s$ in CPL theory) (Lainey et al. 2009; Ferraz-Mello 2012). Then the remaining free parameters would be the satellite's albedo $\alpha_s$ and the orbital eccentricity of the planet-moon orbit $e_{ps}$. $N$-body simulations of the system would allow for an average value of $e_{ps}$.

We conclude that combination of all currently available observational and theoretical techniques can, in principle, yield an estimation of an exomoon's habitability. To that end, the satellite's global average energy flux $\bar{F}_s^{glob}$ (Eq. 22) needs to be compared to the critical flux for a runaway greenhouse $F_{RG}$ (Eq. 1).

## 5. Application to Kepler-22b and KOI211.01

We now apply our stellar-planetary irradiation plus tidal heating model to putative exomoons around Kepler-22b and KOI211.01, both in the habitable zone around their host stars. The former is a confirmed transiting Neptune-sized planet (Borucki et al. 2012), while the latter is a much more massive planet candidate (Borucki et al. 2011). We choose these two planets that likely have very different masses to study the dependence of exomoon habitability on $M_p$.

For the moon, we take our prototype Earth-sized moon and place it in various orbits around the two test planets to investigate a parameter space as broad as possible. We consider two planet-moon semi-major axes: $a_{ps} = 5R_p$, which is similar to Miranda's orbit around Uranus, and $a_{ps} = 20R_p$, which is similar to Titan's orbit around Saturn[11]. We also choose two eccentricities, namely, $e_{ps} = 0.001$ (similar to Miranda) and $e_{ps} = 0.05$ (somewhat larger than Titan's value). With this parametrization, we cover a parameter space of which the diagonal is spanned by Miranda's close, low-eccentricity orbit around Uranus and Titan's far but significantly eccentric orbit around Saturn (see Fig. 8). We must keep in mind, however, that strong additional forces, such as the interaction with further moons, are required to maintain eccentricities of 0.05 in the assumed close orbits for a long time, because tidal dissipation will damp $e_{ps}$. Finally, we consider two orbital inclinations for the moon, $i = 0°$ and $i = 45°$.

[11] Note that in planet-moon binaries close to the star Hill stability requires that the satellite's orbit is a few planetary radii at most. Hence, if they exist, moons about planets in the IHZ around M dwarfs will be close to the planet (Barnes & O'Brien 2002; Domingos et al. 2006; Donnison 2010; Weidner & Horne 2010).

### 5.1 Surface irradiation of putative exomoons about Kepler-22b

Orbiting its solar-mass, solar-luminosity ($T_{eff,*}$ = 5518K, $R_*$ = 0.979$R_\odot$, $R_\odot$ being the radius of the Sun) host star at a distance of ≈0.85AU, Kepler-22b is in the IHZ of a Sun-like star and has a radius 2.38 ± 0.13 times the radius of Earth (Borucki et al. 2012). For the computation of tidal heating on Kepler-22b's moons, we take the parametrization presented in Borucki et al. (2012), and we assume a planetary mass of 25$M_\oplus$, consistent with their photometric and radial-velocity follow-up measurements but not yet well constrained by observations. This ambiguity leaves open the question whether Kepler-22b is a terrestrial, gaseous, or transitional object.

In Fig. 11, we show the photon flux, coming both from the absorbed and re-emitted illumination as well as from tidal heating, at the upper atmosphere for an Earth-sized exomoon in various orbital configurations around Kepler-22b. Illumination is averaged over one orbit of the planet-moon binary around the star. In the upper four panels $a_{ps}$ = 5$R_p$ (similar to the Uranus-Miranda semi-major axis), while in the lower four panels $a_{ps}$ = 20$R_p$ (similar to the Saturn-Titan semi-major axis). The left column shows co-orbital simulations ($i$ = 0°); in the right column $i$ = 45°. In the first and the third line $e_{ps}$ = 0.001; in the second and fourth line $e_{ps}$ = 0.05.

In orbits closer than ≈5$R_p$ even very small eccentricities induce strong tidal heating of exomoons around Kepler-22b. For $e_{ps}$ = 0.001 (upper row), a surface heating flux of roughly 6250W/m$^2$ should induce surface temperatures well above the surface temperatures of Venus, and for $e_{ps}$ = 0.05 (second row), tidal heating is beyond 10$^7$W/m$^2$, probably melting the whole hypothetical moon (Léger et al. 2011). For orbital distances of 20$R_p$, tidal heating is 0.017W/m$^2$ for the low-eccentricity scenario; thus the total flux is determined by stellar irradiation (third line). However, tidal heating is significant for the $e_{ps}$ = 0.05 case (lower line), namely, roughly 42W/m$^2$.

Non-inclined orbits induce strong variations of irradiation over latitude (left column), while for high inclinations, seasons smooth the distribution (right column). As explained in Section 3.1.4, the subplanetary point for co-planar orbits is slightly cooler than the maximum temperature due the eclipses behind the planet once per orbit. But for tilted orbits, the subplanetary point becomes the warmest spot.

We apply the tidal model presented in Heller et al. (2011b) to compute the planet's tilt erosion time $t_{ero}$ and assess whether its primordial obliquity $\psi_p$ could still persist today. Due to its weakly constrained mass, the value of the planet's tidal quality factor $Q_p$ is subject to huge uncertainties. Using a stellar mass of 1$M_\odot$ and trying three values $Q_p$ = 10$^2$, 10$^3$ and 10$^4$, we find $t_{ero}$ = 0.5Gyr, 5Gyr, and 50Gyr, respectively. The lowest $Q_p$ value is similar to that of Earth, while the highest value corresponds approximately to that of Neptune.[12] Thus, if Kepler-22b turns out mostly gaseous and provided that it had a significant primordial obliquity, the planet and its satellites can experience seasons today. But if Kepler-22b is terrestrial and planet-planet perturbations in this system can be neglected, it will have no seasons; and if its moons orbit above the planet's equator, they would share this tilt erosion.

### 5.2 Surface irradiation of putative exomoons about KOI211.01

KOI211.01 is a Saturn- to Jupiter-class planet candidate (Borucki et al. 2011). In the following, we consider it as a planet. Its radius corresponds to 0.88$R_J$, and it has an orbital period of 372.11d around a 6072K main-sequence host star, which yields an estimate for the stellar mass and then a semi-major axis of 1.05AU. The stellar radius is 1.09$R_\odot$, and using models for planet evolution (Fortney et al. 2007), we estimate the planet's mass to be 0.3$M_J$. This value is subject to various uncertainties because little is known about the planet's composition, the mass of a putative core, the planet's atmospheric opacity and structure, and the age of the system. Thus, our investigations will serve as a case study rather than a detailed prediction of exomoon scenarios around this particular planet. Besides KOI211.01, some ten gas giants have been confirmed in the IHZ of their host stars, all of which are not transiting. Thus, detection of their putative moons will not be feasible in the foreseeable future.

In Fig. 12, we present the flux distribution on our prototype Earth-like exomoon around KOI211.01 in the same orbital configurations as in the previous subsection. Tidal heating in orbits with $a_{ps} \lesssim 5R_p$ can be strong, depending on eccentricity, and make the moon uninhabitable. For $e_{ps} = 0.001$, tidal heating is of the order of 1W/m$^2$ and thereby almost negligible for the moderate flux distribution (first line). However, it increases to over 3200W/m$^2$ for $e_{ps} = 0.05$ and makes the moon uninhabitable (second line). Tidal heating is negligible at semi-major axes $\gtrsim 20R_p$ even for a significant eccentricity of 0.05.

[12] See Gavrilov & Zharkov (1977) and Heller et al. (2010) for discussions of $Q$ values and Love numbers for gaseous substellar objects.

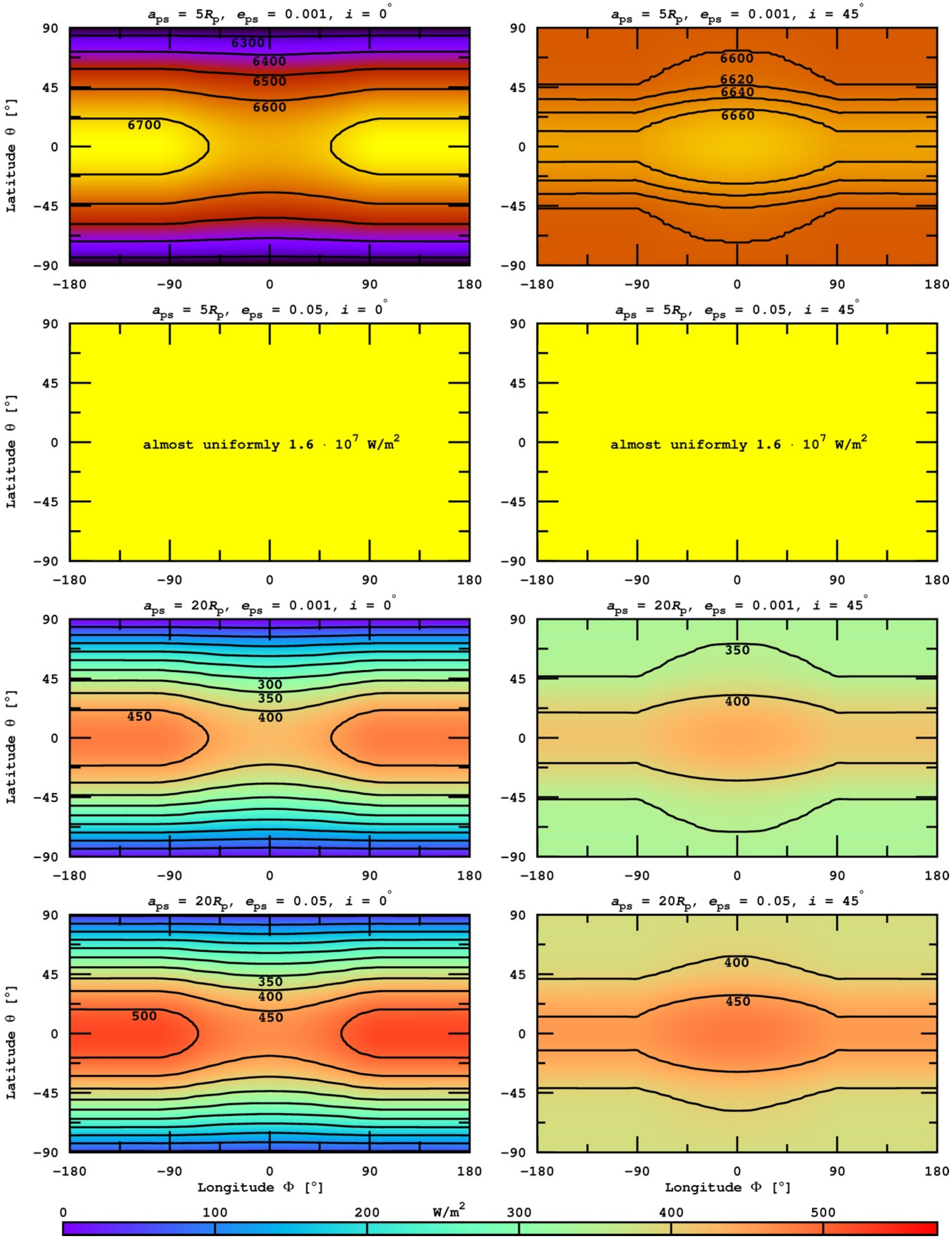

**FIG. 11.** Orbit-averaged flux (in units of W/m²) at the top of an Earth-sized exomoon's atmosphere around Kepler-22b for eight different orbital configurations. Computations include irradiation from the star and the planet as well as tidal heating. The color bar refers only to the lower two rows with moderate flux.

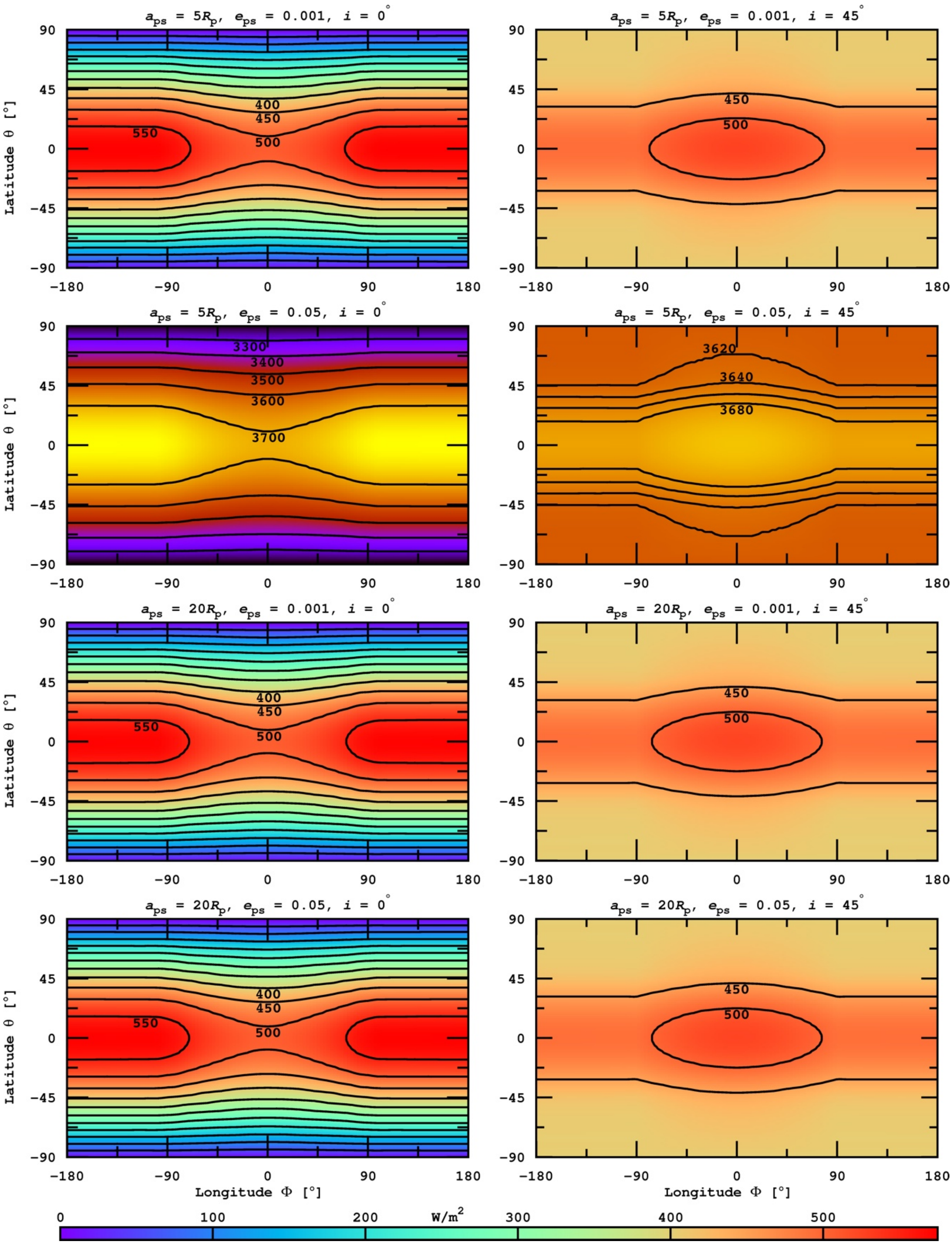

**FIG. 12.** Orbit-averaged flux (in units of W/m²) at the top of an Earth-sized exomoon's atmosphere around KOI211.01 for eight different orbital configurations. Computations include irradiation from the star and the planet as well as tidal heating. The color bar refers only to the first and the lower two rows with moderate flux.

By comparison of Fig. 11 with Fig. 12, we see that in those cases where tidal heating can be neglected and stellar irradiation dominates the energy flux, namely, in the third line of both figures, irradiation is higher on comparable moons around KOI211.01. But in the lower panel line, the total flux on our Kepler-22b prototype satellite becomes comparable to the one around KOI211.01, which is because of the extra heat from tidal dissipation. Surface flux distributions in the low-eccentricity and the moderate-eccentricity state on our KOI211.01 exomoons are virtually the same at $a_{ps} = 20R_p$.

Since KOI211.01 is a gaseous object, we apply a tidal quality factor of $Q_p = 10^5$, which is similar to, but still lower than, the tidal response of Jupiter. We find that $t_{ero}$ of KOI211.01 is much higher than the age of the Universe. Thus, satellites of KOI211.01 will experience seasons, provided the planet had a primordial obliquity and planet-planet perturbations can be neglected.

The radius of KOI211.01 is about 4 times greater than the radius of Kepler-22b; thus moons at a certain multitude of planetary radii distance from the planet, say $5R_p$ or $20R_p$ as we considered, will be effectively much farther away from KOI211.01 than from Kepler-22b. As tidal heating strongly depends on $a_{ps}$ and on the planet's mass, a quick comparison between tidal heating on exomoons around Kepler-22b and KOI211.01 can be helpful. For equal eccentricities, the fraction of tidal heating in two moons about Kepler-22b and KOI211.01 will be equal to a fraction of $Z_s$ from Eq. (20). By taking the planet masses assumed above, thus $M_{KOI} \approx 3.81M_{Ke}$, and assuming that the satellite mass is much smaller than the planets' masses, respectively, we deduce

$$\frac{Z_s^{Ke}}{Z_s^{KOI}} = \frac{M_{Ke}^2}{M_{KOI}^2} \frac{(M_{Ke} + M_s)}{(M_{KIO} + M_s)} \left( \frac{a_{ps}^{KOI}}{a_{ps}^{Ke}} \right)^9 \overset{M_s \text{ negligible}}{\approx} \frac{4^9}{3.81^3} \approx 5000 \ , \quad (23)$$

where indices 'Ke' and 'KOI' refer to Kepler-22b and KOI211.01, respectively. The translation of tidal heating from any panel in Fig. 11 to the corresponding panel in Fig. 12 can be done with a division by this factor.

### 5.3 Orbits of habitable exomoons

We now set our results in context with observables to obtain a first estimate for the magnitude of the TTV amplitude induced by habitable moons around Kepler-22b and KOI211.01. For the time being, we will neglect the actual detectability of such signals with *Kepler* but discuss it in Section 6 (see also Kipping et al. 2009).

To begin with, we apply our method from Section 4 and add the orbit-averaged stellar irradiation on the moon to the averaged stellar-reflected light, the averaged thermal irradiation from the planet, and the tidal heating (in the CTL theory). We then compare their sum $\bar{F}_s^{glob}$ (see Eq. 22) to the critical flux for a runaway greenhouse $F_{RG}$ on the respective moon. In Fig. 13, we show the limiting orbits ($\bar{F}_s^{glob} = F_{RG}$) for a runaway greenhouse of an Earth-like (upper row) and a Super-Ganymede (lower row) exomoon around Kepler-22b (left column) and KOI211.01 (right column). For both moons, we consider two albedos ($\alpha_s = 0.3$, gray solid lines; and $\alpha_s = 0.4$, black solid lines) and three orbital eccentricities ($e_{ps} = 0.001$, $0.01, 0.1$). Solid lines define the habitable edge as described in Section 4. Moons in orbits to the left of a habitable edge are uninhabitable, respectively, because the sum of stellar and planetary irradiation plus tidal heating exceeds the runaway greenhouse limit. Dashed lines correspond to TTV amplitudes and will be discussed below.

#### 5.3.1 Orbit-averaged global flux

Let us first consider the three black solid lines in the upper left panel, corresponding to an Earth-like satellite with $\alpha_s = 0.4$ about Kepler-22b. Each contour represents a habitable edge for a certain orbital eccentricity of the moon; that is, in these orbits the average energy flux on the moon is equal to $F_{RG} = 295$W/m$^2$. Assuming a circular orbit around the star ($e_{*p} = 0$), the orbit-averaged stellar flux absorbed by a moon with $\alpha_s = 0.4$ is

$$\frac{L_*(1 - \alpha_s)}{16\pi a_{*p}^2} \approx 227\,\mathrm{W/m^2} \ , \quad (29)$$

using the parametrization presented in Section 5.1. Thus, black contours indicate an additional heating of $\approx$68W/m$^2$. The indicated eccentricities increase from left to right. This is mainly due to tidal heating, which increases for larger eccentrici-ties but decreases at larger separations $a_{ps}$. Thus, the larger $e_{ps}$, the farther away from the planet the moon needs to be to avoid becoming a greenhouse. If a moon is close to the planet and shall be habitable, then its eccentricity must be small enough.

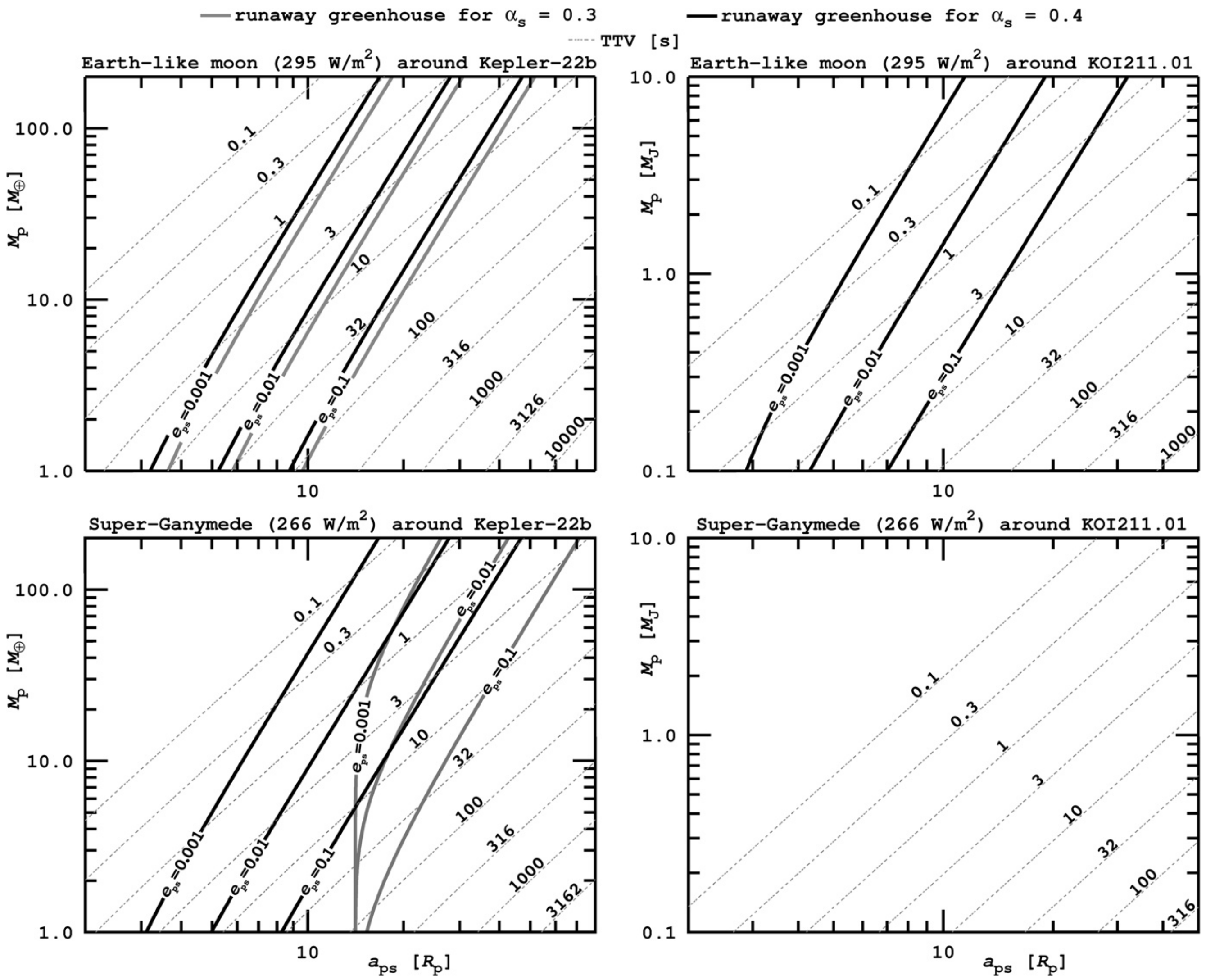

**FIG. 13.** Habitable edges for an Earth-like (upper row) and a Super-Ganymede (lower row) exomoon in orbit around Kepler-22b (left column) and KOI211.01 (right column). Masses of both host planets are not well constrained; thus abscissae run over several decades (in units of $M_\oplus$ for Kepler-22b and $M_J$ for KOI211.01). We consider two albedos $\alpha_s = 0.3$ (gray solid lines) and $\alpha_s = 0.4$ (black solid lines) and three eccentricities ($e_{ps} = 0.001, 0.01, 0.1$) for both moons. No gray solid lines are in the right column because both prototype moons would not be habitable with $\alpha_s = 0.3$ around KOI211.01. TTV amplitudes (in units of seconds) for coplanar orbits are plotted with dashed lines.

Next, we compare the black lines to the gray lines, which assume an albedo of 0.3 for an Earth-like satellite. With this albedo, it absorbs ≈265W/m$^2$ of stellar irradiation. Consequently, a smaller amount of additional heat is required to push it into a runaway greenhouse, so the corresponding habitable edges are farther away from the planet, where both tidal heating and irradiation from the planet are lower.

In the lower left panel, we consider a Super-Ganymede around Kepler-22b. Its critical flux of 266W/m$^2$ is smaller than for an Earth-like moon, so for an albedo of 0.4 only 39W/m$^2$ of additional heating is required for the moon to turn into a runaway greenhouse. Yet, compared to the Earth-like satellite in the upper left panel, the habitable edges (black lines) are still slightly closer to the planet, because tidal heating in the smaller Super-Ganymede is much weaker at a given orbital-semi major axis. More important, gray lines are at much larger distances in this panel because with an albedo of 0.3 the satellite's absorbed stellar flux (265W/m$^2$) is almost as high as the critical flux (266W/m$^2$). This is irrespective of whether our Ganymede-like object is orbiting a planet.

Next, we consider moons in orbit about KOI211.01, shown in the right column of Fig. 13. First note the absence of gray solid lines. For moons with an albedo of $\alpha_s = 0.3$, the orbit-averaged stellar irradiation will be 315W/m$^2$, with the parametrization presented in Sect 5.2. This means that stellar irradiation alone is larger than the critical flux of both our Earth-like (295W/m$^2$, upper panel) and Super-Ganymede (266W/m$^2$, lower panel) prototype moons. Thus, both moons with $\alpha_s = 0.3$ around KOI211.01 are not habitable, irrespective of their distance to the planet.

Moons with $\alpha_s = 0.4$ around KOI211.01 absorb 270W/m$^2$, which is less than the critical flux for the Earth-like moon (upper right panel) but more than the Super-Ganymede moon could bear (lower right panel). Thus, the black lines are absent from the latter plot but show up in the upper right. Contours for given eccentricities are closer to the planet than their counterparts in the case of Kepler-22b. This is because we plot the semi-major axis in units of planetary radii – while KOI211.01 has a much larger radius than Kepler-22b – and tidal heating, which strongly depends on the *absolute* distance between the planet and the moon.

#### 5.3.2 Transit timing variations of habitable exomoons

Now we compare the habitable edges as defined by Eq. (22) to the amplitude of the TTV induced by the moon on the planet. TTV amplitudes $\Delta$ in units of seconds are plotted with dashed lines in Fig. 13. Recall, however, that $\Delta$ would not directly be measured by transit observations. Rather the root-mean-square of the TTV wave would be observed (Kipping 2009a). We apply the Kipping et al. (2012) equations, assuming that the moon's orbit is circular and that both the circumstellar and the circumplanetary orbit are seen edge-on from Earth.

Each panel of Fig. 13 shows how the TTV amplitude decreases with decreasing semi-major axis of the satellite (from right to left) and with increasing planetary mass (from bottom to top). Thus, for a given planetary mass, a moon could be habitable if its TTV signal is sufficiently large. Comparison of the upper and the lower panels in each column shows how much larger the TTV amplitude of an Earth-like moon is with respect to our Super-Ganymede moon.

Corrections due to an inclination of the moon's orbit and an accidental alignment of the moon's longitude of the periapses are not included but would be small in most cases. For non-zero inclinations, the TTV amplitude $\Delta$ will be decreased. This decrease is proportional to $\sqrt{1 - \cos(i_s)^2 \sin(\varpi_s)^2}$ , where $i_s$ is measured from the circumstellar orbital plane normal to the orbital plane of the planet-moon orbit, and $\varpi_s$ denotes the orientation of the longitude of the periapses (see Eq. (6.46) in Kipping 2011b). Thus, corrections to our picture will only be relevant if both the moon's orbit is significantly tilted and $\varpi_s \approx 90°$. In the worst case of $\varpi_s \approx 90°$, an inclination of 25° reduces $\Delta$ by <10%, while for $\varpi_s \approx 45°$, $i_s$ could be as large as 40° to produce a similar correction. Simulations of the planet's and the satellite's tilt erosion can help assess whether substantial misalignments are likely (Heller et al. 2011b). While such geometric blurring should be small for most systems, prediction of the individual TTV signal for a satellite in an exoplanet system – as we suggest in Fig. 13 – is hard because perturbations of other planets, moons, or Trojans could affect the TTV.

## 6. Summary and discussion

Our work yields the first translation from observables to exomoon habitability. Using a scaling relation for the onset of the runaway greenhouse effect, we have deduced constraints on exomoon habitability from stellar and planetary irradiation as well as from tidal heating. We determined the orbit-averaged global energy budget $\bar{F}_s^{\rm glob}$ for exomoons to avoid a runaway greenhouse and found that for a well parametrized system of a star, a host planet, and a moon, Eq. (22) can be used to evaluate the habitability of a moon. By analogy with the circumstellar habitable zone (Kasting et al. 1993), these rules define a circumplanetary "habitable edge". To be habitable, moons must orbit their planets outside the habitable edge.

Application of our illumination plus tidal heating model shows that an Earth-sized exomoon about Kepler-22b with a bond albedo of 0.3 or higher would be habitable if (*i.*) the planet's mass is $\approx 10 M_\oplus$, (*ii.*) the satellite would orbit Kepler-22b with a semi-major axis $\gtrsim 10 R_p$, and (*iii.*) the moon's orbital eccentricity $e_{ps}$ would be <0.01 (see Fig. 13). If Kepler-22b turns out more massive, then such a putative moon would need to be farther away to be habitable. Super-Ganymede satellites of Kepler-22b meet similar requirements, but beyond that their bond albedo needs to be $\gtrsim 0.4$.

Super-Ganymede or smaller moons around KOI211.01 will not be habitable, since their critical flux for a runaway greenhouse effect ($\lesssim$266W/m$^2$) is less than the orbit-averaged irradiation received by the star (270W/m$^2$). But Earth-like or more massive moons can be habitable if (*i.*) the planet's mass $\approx M_J$, (*ii.*) the satellite's bond albedo $\gtrsim 0.4$, (*iii.*) the satellite orbits KOI211.01 with a semi-major axis $\gtrsim 10 R_p$, and (*iv.*) the moon's orbital eccentricity $e_{ps} < 0.01$. If the planet turns out less massive, then its moons could be closer and have higher eccentricities and still be habitable.

Kepler-22 and KOI211 both are mid-G-type stars, with *Kepler* magnitudes 11.664 (Borucki et al. 2012) and 14.99 (Borucki et al. 2011), respectively. As shown by Kipping et al. (2009), the maximum *Kepler* magnitude to allow for the detection of an Earth-like moon is about 12.5. We conclude that such moons around Kepler-22b are detectable if they exist, and their habitability could be evaluated with the methods provided in this communication. Moons around KOI211.01 will not be detectable within the 7 year duty cycle of *Kepler*. Nevertheless, our investigation of this giant planet's putative moons serves as a case study for comparably massive planets in the IHZ of their parent stars.

Stellar flux on potentially habitable exomoons is much stronger than the contribution from the planet. Nevertheless, the sum of thermal emission and stellar reflected light from the planet can have a significant impact on exomoon climates. Stellar reflection dominates over thermal emission as long as the planet's bond albedo $\alpha_p \gtrsim 0.1$ and can reach some 10 or even 100W/m$^2$ once the moon is in a close orbit ($a_{ps} \lesssim 10R_p$) and the planet has a high albedo ($\alpha_p \gtrsim 0.3$) (Section 3.1.2). Precise values depend on stellar luminosity and the distance of the planet-moon duet from the star. Our calculations for Kepler-22b and KOI211.01 show that the limiting orbits for exomoons to be habitable are very sensitive to the satellite's albedo.

Due to the weak tides from its host star, KOI211.01 can still have a significant obliquity, and if its moons orbit the planet in the equatorial plane they could have seasons and are more likely to be discovered by transit duration variations of the transit impact parameter (TDV-TIP, Kipping 2009b). For Kepler-22b, the issue of tilt erosion cannot be answered unambiguously until more about the planet's mass and composition is known.

If a moon's orbital inclination is small enough, then it will be in the shadow of the planet for a certain time once per planet-moon orbit (see also Heller 2012). For low inclinations, eclipses can occur about once every revolution of the moon around the planet, preferentially when the subplanetary hemisphere on the moon would experience stellar irradiation maximum (depending on $e_{*p}$). Eclipses have a profound impact on the surface distribution of the moon's irradiation. For low inclinations, the subplanetary point on the moon will be the "coldest" location along the equator, whereas for moderate inclinations it will be the "warmest" spot on the moon due to the additional irradiation from the planet. Future investigations will clarify whether this may result in enhanced sub*planetary* weathering instabilities, that is, runaway $CO_2$ drawdown rates eventually leading to very strong greenhouse forcing, or sub*planetary* dissolution feedbacks of volatiles in sub*planetary* oceans, as has been proposed for exoplanets that are tidally locked to their host stars and thus experience such effects at the fixed sub*stellar* point (Kite et al. 2011).

We predict seasonal illumination phenomena on the moon, which emerge from the circumstellar season and planetary illumination. They depend on the location on the satellite and appear in four versions, which we call the "proplanetary summer", "proplanetary winter", "antiplanetary summer", and "antiplanetary winter". The former two describe seasons due to the moon's obliquity with respect to the star with an additional illumination from the planet; the latter two depict the permanent absence of planetary illumination during the seasons.

For massive exomoons with $a_{ps} \lesssim 10R_p$ around Kepler-22b and around KOI211.01, tidal heating can be immense, presumably making them uninhabitable if the orbits are substantially non-circular. On the one hand, tidal heating can be a threat to life on exomoons, in particular when they are in close orbits with significant eccentricities around their planets. If the planet-moon duet is at the inner edge of the circumstellar IHZ, small contributions of tidal heat can render an exomoon uninhabitable. Tidal heating can also induce a thermal runaway, producing intense magmatism and rapid resurfacing on the moon (Běhounková et al. 2011). On the other hand, we can imagine scenarios where a moon becomes habitable only because of tidal heating. If the host planet has an obliquity similar to Uranus, then one polar region will not be illuminated for half the orbit around the star. Moderate tidal heating of some tens of watts per square meter might be just adequate to prevent the atmosphere from freezing out. Or if the planet and its moon orbit their host star somewhat beyond the outer edge of the IHZ, then tidal heating might be necessary to make the moon habitable in the first place. Tidal heating could also drive long-lived plate tectonics, thereby enhancing the moon's habitability (Jackson et al. 2008). An example is given by Jupiter's moon Europa, where insolation is weak but tides provide enough heat to sustain a subsurface ocean of liquid water (Greenberg et al. 1998; Schmidt et al. 2011). On the downside, too much tidal heating can render the body uninhabitable due to enhanced volcanic activity, as it is observed on Io.

Tidal heating has a strong dependence on the moon's eccentricity. Eccentricities of exomoons will hardly be measurable even with telescopes available in the next decade, but it will be possible to constrain $e_{ps}$ by simulations. Therefore, once exomoon systems are discovered, it will be necessary to search for further moons around the same planet to consistently simulate the $N$-body ($N > 2$) evolution with multiple-moon interaction, gravitational perturbations from other planets, and the gravitational effects of the star. Such simulations will also be necessary to simulate the long-term evolution of the orientation $\eta$ of the moon's inclination $i$ with respect to the periastron of the star-related orbit, because for signifiant eccentricities $e_{*p}$ it will make a big difference whether the summer of either the northern or southern hemisphere coincides with minimum or maximum distance to the star. Technically, these variations refer to the apsidal precession (the orientation of the star-related eccentricity, i.e., of $a_{*p}$) and the precession of the planet's rotation axis, both of which determine $\eta$. For Saturn, and thus Titan, the corresponding time scale is of order 1Myr (French et al. 1993), mainly induced by solar torques on both Saturn's oblate figure and the equatorial satellites. Habitable exomoons might preferably be irregular satellites (see Section 2) for which Carruba et al. (2002) showed that their orbital parameters are subject to particularly rapid changes, driven by stellar perturbations.

We find that more massive moons can orbit more closely to the planet and be prevented from becoming a runaway greenhouse (Section 2.2). This purely atmospheric effect is shared by all terrestrial bodies. Similar to the circumstellar habitable zone of extrasolar planets (Kasting et al. 1993), we conclude that more massive exomoons may have somewhat wider habitable zones around their host planets – of which the inner boundary is defined by the habitable edge and the outer boundary by Hill stability – than do less massive satellites. In future investigations, it will be necessary to include simulations of the moons' putative atmospheres and their responses to irradiation and tidal heating. Thus, our irradiation plus tidal heating model should be coupled to an energy balance or global climate model to allow for more realistic descriptions of exomoon habitability. As indicated by our basal considerations, the impact of eclipses and planetary irradiation on exomoon climates can be substantial. In addition to the orbital parameters which we have simulated here, the moons' climates will depend on a myriad bodily characteristics

Spectroscopic signatures of life, so-called "biosignatures", in the atmospheres of inhabited exomoons will only be detectable with next-generation, several-meter-class space telescopes (Kaltenegger 2010; Kipping et al. 2010). Until then, we may primarily use our knowledge about the orbital configurations and composition of those worlds when assessing their habitability. Our method allows for an evaluation of exomoon habitability based on the data available at the time they will be discovered. The recent detection of an Earth-sized and a sub-Earth-sized planet around a G-type star (Fressin et al. 2012) suggests that not only the moons' masses and semi-major axes around their planets can be measured (e.g., by combined TTV and TDV, Kipping 2009a) but also their radii by direct photometry. A combination of these techniques might finally pin down the moon's inclination (Kipping 2009b) and thus allow for precise modeling of its habitability based on the model presented here.

Results of ESA's *Jupiter Icy Moons Explorer* ("*JUICE*") will be of great value for characterization of exomoons. With launch in 2022 and arrival at Jupiter in 2030, one of the mission's two key goals will be to explore Ganymede, Europa, and Callisto as possible habitats. Therefore, the probe will acquire precise measurements of their topographic distortions due to tides on a centimeter level; determine their dynamical rotation states (i.e., forced libation and nutation); characterize their surface chemistry; and study their cores, rocky mantles, and icy shells. The search for water reservoirs on Europa, exploration of Ganymede's magnetic field, and monitoring of Io's volcanic activity will deliver fundamentally new insights into the planetology of massive moons.

Although our assumptions about the moons' orbital characteristics are moderate, that is, they are taken from the parameter space mainly occupied by the most massive satellites in the Solar System, our results imply that exomoons might exist in various habitable or extremely tidally heated configurations. We conclude that the advent of exomoon observations and characterization will permit new insights into planetary physics and reveal so far unknown phenomena, analogous to the staggering impact of the first exoplanet observations 17 years ago. If observers feel animated to use the available *Kepler* data, the *Hubble Space Telescope,* or meter-sized ground-based instruments to search for evidence of exomoons, then one aim of this communication has been achieved.

## Appendix

### Appendix A: Stellar irradiation

We include here a thorough explanation for the stellar flux $f_*(t)$ presented in Section 3.1.1. To begin with, we assume that the irradiation on the moon at a longitude $\phi$ and latitude $\theta$ will be

$$f_*(t) = \frac{L_*}{4\pi \vec{r}_{s*}(t)^2} \frac{\vec{r}_{s*}(t)}{r_{s*}(t)} \frac{\vec{n}_{\phi,\theta}(t)}{n_{\phi,\theta}(t)} \ , \qquad (A.1)$$

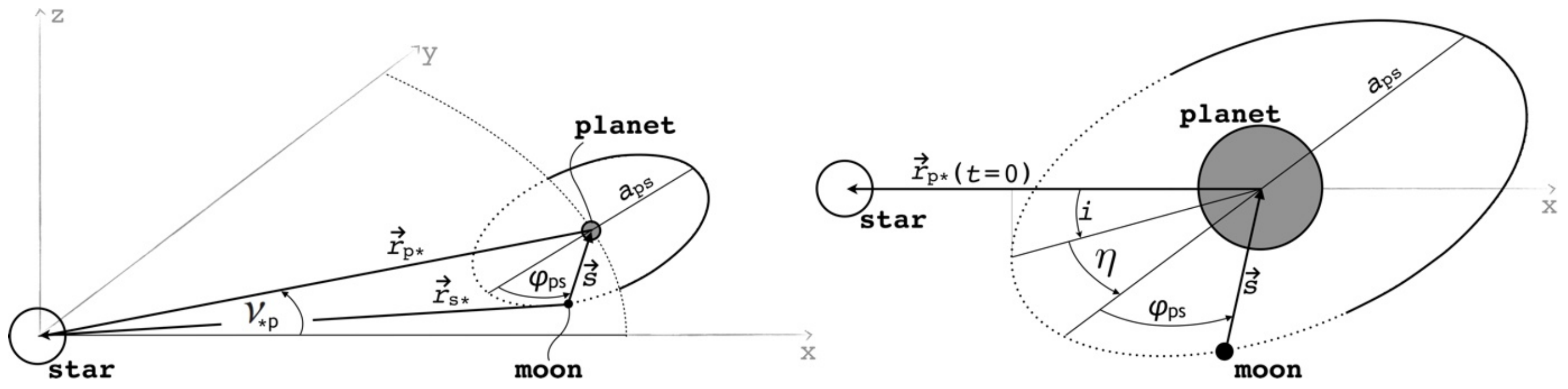

**FIG. A.1.** Geometry of the triple system of a star, a planet, and a moon. In the left panel the planet-moon duet has advanced by an angle $\nu_{*p}$ around the star, and the moon has progressed by an angle $2\pi\varphi_{ps}$. The right panel shows a zoom-in to the planet-moon binary. As in the left panel, time has proceeded, and a projection of $\vec{r}_{p*}(t)$ at time $t = 0$ has been included to explain the orientation of the moon's orbit, which is inclined by an angle $i$ and rotated against the star-moon periapses the by an angle $\eta$.

with $L_*$ as the stellar luminosity and $\vec{r}_{s*}(t)$ as the vector from the moon to the star (see Fig. A.1). The product of the surface normal $\vec{n}_{\phi,\theta}/n_{\phi,\theta}$ on the moon and $\vec{r}_{s*}(t)/r_{s*}(t)$ accounts for projection effects on the location $(\phi,\theta)$, and we set $f_*$ to zero in those cases where the star shines from the back. Figure A.1 shows that $\vec{r}_{s*}(t) = \vec{r}_{p*}(t) + \vec{s}(t)$. While the Keplerian motion, encapsulated in $\vec{r}_{p*}(t)$, is deduced in Section 3.1.1, we focus here on the moon's surface normal $\vec{n}_{\phi,\theta}$.

In the right panel of Fig. A.1, we show a close-up of the star-planet-moon geometry at $t = 0$, which corresponds to the initial configuration of the orbit evolution. The planet is at periastron ($\varphi_{*p} = 0$) and for the case where $\eta = 0$ the vector $\vec{s} = \vec{n}_{0,0}(t)$ from the subplanetary point on the moon to the planet would have the form

$$a_{ps} \begin{pmatrix} \cos\left(2\pi\frac{(t-\tau)}{P_{ps}}\right)\cos\left(i\frac{\pi}{180°}\right) \\ \sin\left(2\pi\frac{(t-\tau)}{P_{ps}}\right) \\ \cos\left(2\pi\frac{(t-\tau)}{P_{ps}}\right)\sin\left(i\frac{\pi}{180°}\right) \end{pmatrix}, \qquad \text{(A.2)}$$

where all angles are provided in degrees. The angle $\eta$ depicts the orientation of the lowest point of the moon's orbit with respect to the projection of $\vec{r}_{p*}$ on the moon's orbital plane at $t = 0$. At summer solstice on the moon's nothern hemisphere, the true anomaly $\nu_{*p}$ equals $\eta$ (see Eq. B.8). This makes $\eta$ a critical parameter for the seasonal variation of stellar irradiation on the moon because it determines how seasons, induced by orbital inclination $i$, relate to the changing distance to the star, induced by star-planet eccentricity $e_{*p}$. In particular, if $\eta = 0$ then northern summer coincides with the periastron passage about the star and nothern winter occurs at apastron, inducing distinctly hot summers and cold winters. It relates to the conventional orientation of the ascending node $\Omega$ as $\eta = \Omega + 270°$ and can be considered as the climate-precession parameter.

For $\eta \neq 0$, we have to apply a rotation $M(\eta)$: $\mathbb{R}^3 \rightarrow \mathbb{R}^3$ of $\vec{s}$ around the $z$-axis (0,0,1), which is performed by the rotation matrix

$$M(\eta) = \begin{pmatrix} \cos(\eta) & -\sin(\eta) & 0 \\ \sin(\eta) & \cos(\eta) & 0 \\ 0 & 0 & 1 \end{pmatrix}. \qquad \text{(A.3)}$$

With the abbreviations introduced in Eq. (6) (with $\phi = 0$ for the time being), we obtain

$$\vec{s}(t) \equiv \vec{n}_{0,0}(t) = a_{ps} \begin{pmatrix} \tilde{C}cC - \tilde{S}s \\ \tilde{S}cC - \tilde{C}s \\ cS \end{pmatrix}. \qquad \text{(A.4)}$$

This allows us to parametrize the surface normal $\vec{s}(t)/s(t)$ of the subplanetary point on the moon for arbitrary $i$ and $\eta$.

Finally, we want to find the surface normal $\vec{n}_{\phi,\theta}(t)/n_{\phi,\theta}(t)$ for *any* location ($\phi$,$\theta$) on the moon's surface. Therefore, we have to go two steps on the moon's surface, each one parametrized by one angle: one in longitudinal direction $\phi$ and one in latitudinal direction $\theta$. The former one can be walked easily, just by adding $2\pi\phi/360°$ in the sine and cosine arguments in Eq. (A.2) (see first line in Eq. 6), thus $\vec{n}_{\phi,0} = \vec{s}(\varphi_{\rm ps} + 2\pi\phi/360°)$. This is equivalent to a shift along the moon's equator. For the second step, we know that if $\theta = 90°$, then the surface normal will be along the rotation axis

$$\vec{N} = a_{\rm ps} \begin{pmatrix} -S\tilde{C} \\ -S\tilde{S} \\ C \end{pmatrix} \qquad \text{(A.5)}$$

of the satellite; that is, we are standing on the north pole. The vector $\vec{n}_{\phi,\theta}$ can then be obtained by tilting $\vec{s}(\varphi_{\rm ps} + 2\pi\phi/360°)$ by an angle $\theta$ toward $\vec{N}$. With $N = a_{\rm ps} = s$, we then derive

$$\begin{aligned}\vec{n}_{\phi,\theta}(t) &= a_{\rm ps}\sin(\theta)\frac{\vec{N}}{N} + a_{\rm ps}\cos(\theta)\frac{\vec{s}(\varphi_{\rm ps} + 2\pi\phi/360°)}{s(\varphi_{\rm ps} + 2\pi\phi/360°)} = \sin(\theta)\vec{N} + \cos(\theta)\vec{s}(\varphi_{\rm ps} + 2\pi\phi/360°) \\ &= a_{\rm ps}\begin{pmatrix} -\bar{s}S\tilde{C} + \bar{c}(\tilde{C}cC - \tilde{S}s) \\ -\bar{s}S\tilde{S} + \bar{c}(\tilde{S}cC - \tilde{C}s) \\ \bar{s}C + \bar{c}cS \end{pmatrix}, \qquad \text{(A.6)}\end{aligned}$$

which solves Eq. (A.1).

## Appendix B: Planetary irradiation

In addition to stellar light, the moon will receive thermal and stellar-reflected irradiation from the planet, $f_{\rm t}(t)$ and $f_{\rm r}(t)$, respectively. One hemisphere on the planet will be illuminated by the star and the other one will be dark. On the bright side of the planet, there will be both thermal emission from the planet as well as starlight reflection, while on the dark side the planet will only emit thermal radiation, though with a lower intensity than on the bright side due to the lower temperature.

We begin with the thermal part. The planet's total thermal luminosity $L_{\rm th,p}$ will be the sum of the radiation from the bright side and from the dark side. On the bright side, the planet shall have a uniform temperature $T^{\rm b}_{\rm eff,p}$, and on the dark side its temperature shall be $T^{\rm d}_{\rm eff,p}$. Then thermal equilibrium between outgoing thermal radiation and incoming stellar radiation yields

$$\begin{aligned}L_{\rm th,p} &= 2\pi R_{\rm p}^2\sigma_{\rm SB}\Big((T^{\rm b}_{\rm eff,p})^4 + (T^{\rm d}_{\rm eff,p})^4\Big) \\ &= \pi R_{\rm p}^2(1-\alpha_{\rm p})\frac{4\pi R_*^2\sigma_{\rm SB}T^4_{\rm eff,*}}{4\pi\vec{r}^{\,2}_{\rm p*}} + W_{\rm p}, \qquad \text{(B.1)}\end{aligned}$$

where the first term in the second line describes the absorbed radiation from the star and the second term ($W_{\rm p}$) can be any additional heat source, for example, the energy released by the gravitation-induced shrinking of the gaseous planet. For Jupiter, Saturn, and Neptune, which orbit the Sun at distances >5AU, $W_{\rm p}$ is greater than the incoming radiation. However, for gaseous planets in the IHZ it will be negligible once the planet has reached an age ≳ 100Myr (Baraffe et al. 2003).[13] Owed to the negligibility for our purpose and for simplicity we set $W_{\rm p} = 0$.

In our model, we want to parametrize the two hemispheres by a temperature difference $\mathrm{d}T \equiv T^{\rm b}_{\rm eff,p} - T^{\rm d}_{\rm eff,p}$. We define

$$p(T^{\rm b}_{\rm eff,p}) \equiv (T^{\rm b}_{\rm eff,p})^4 + (T^{\rm b}_{\rm eff,p} - \mathrm{d}T)^4 - T^4_{\rm eff,*}\frac{(1-\alpha_{\rm p})R_*^2}{2\vec{r}^{\,2}_{\rm *p}} = 0 \qquad \text{(B.2)}$$

[13] Note that Barnes et al. (2013) identified 100Myr as the time required for a runaway greenhouse to sterilize a planet.

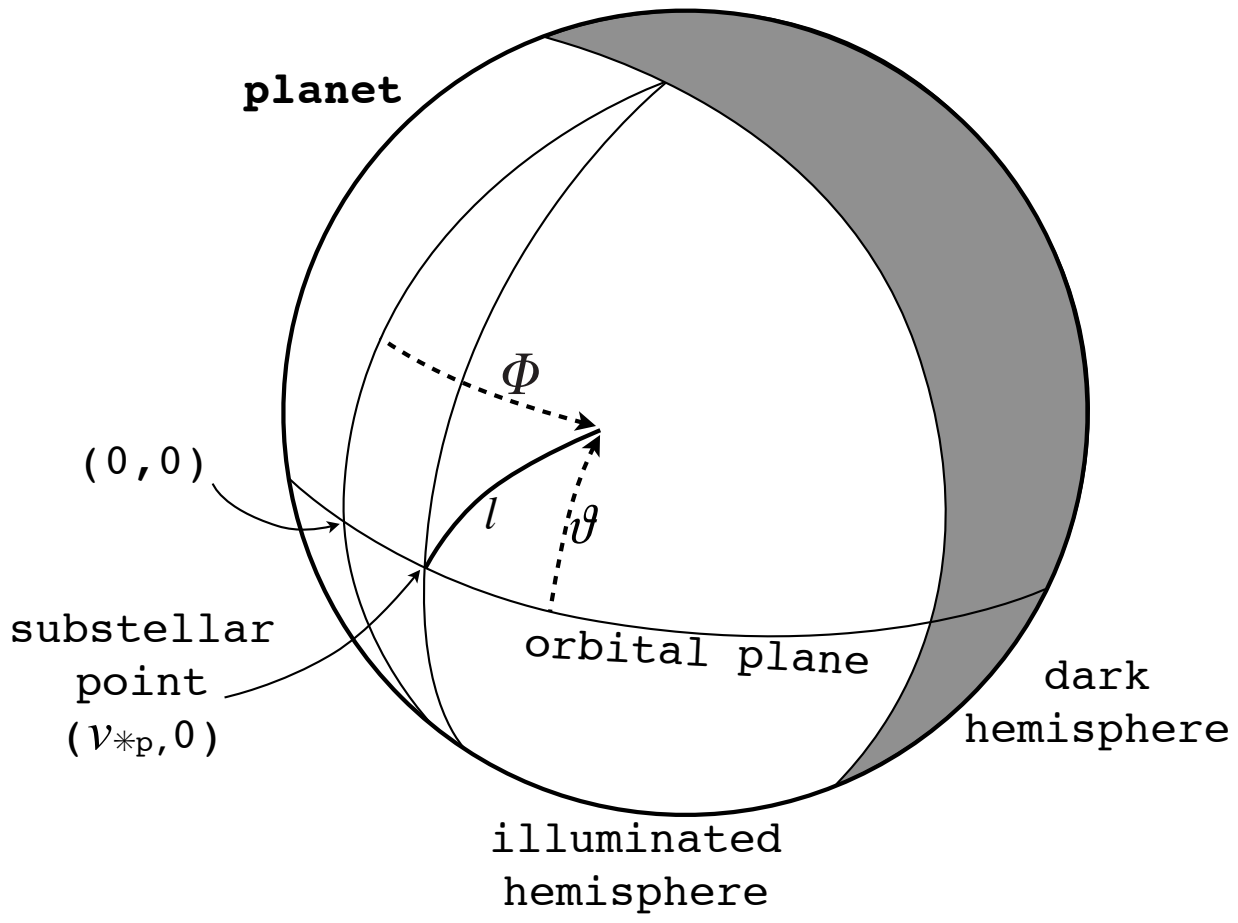

**FIG. B.1.** Geometry of the planetary illumination. The sub-satellite point on the planetary surface is at ($\Phi$,$\vartheta$). The angular distance $l$ of the moon from the substellar point ($\nu_{*p}$,0) determines the amount of light received by the moon from the two different hemispheres.

and, once d$T$ is given, search for the first zero point of $p(T_{\rm eff,p}^{\rm b})$ above

$$T_{\rm eff,p}^{\rm eq} = T_{\rm eff,*} \left( \frac{(1-\alpha_{\rm p})R_*^2}{4\vec{r}_{*\rm p}^{\,2}} \right)^{1/4} \ , \qquad \text{(B.3)}$$

which yields $T_{\rm eff,p}^{\rm b}$ and $T_{\rm eff,p}^{\rm d}$. In our prototype exomoon system, we assume d$T$ = 100K, which is equivalent to fixing the efficiency of heat redistribution from the bright to the dark hemisphere on the planet (Burrows et al. 2006b; Budaj et al. 2012).

From Eq. (B.1), we can deduce the thermal flux on the subplanetary point on the moon. With increasing angular distance from the subplanetary point, $f_{\rm t}$ will decrease. We can parametrize this distance on the moon's surface by longitude $\phi$ and latitude $\theta$, which gives

$$\begin{aligned} f_{\rm t}(t) \ &= \frac{R_{\rm p}^2 \sigma_{\rm SB}}{a_{\rm ps}^2} \cos\left(\frac{\phi\pi}{180^\circ}\right) \cos\left(\frac{\theta\pi}{180^\circ}\right) \\ &\times \left[ (T_{\rm eff,p}^{\rm b})^4 \xi(t) + (T_{\rm eff,p}^{\rm d})^4 \Big(1-\xi(t)\Big) \right] \ . \end{aligned} \qquad \text{(B.4)}$$

The time dependence of the irradiation is now packed into $\xi(t)$, which serves as a weighting function for the two contributions from the bright and the dark side. It is given by

$$\xi(t) = \frac{1}{2}\left(1 + \cos\Big(l(t)\Big)\right) \ , \qquad \text{(B.5)}$$

where

$$l(t) = \arccos\left\{ \cos\Big(\vartheta(t)\Big) \cos\Big(\Phi(t) - \nu_{*\rm p}(t)\Big) \right\} \qquad \text{(B.6)}$$

is the angular distance between the moon's projection on the planetary surface and the substellar point on the planet. In other words, $l(t)$ is an orthodrome on the planet's surface, determined by $\Phi(t)$ and $\vartheta(t)$ (see Fig. B.1). This yields

$$\xi(t) = \frac{1}{2}\left\{1 + \cos\Big(\vartheta(t)\Big) \cos\Big(\Phi(t) - \nu_{*\rm p}(t)\Big) \right\} \ . \qquad \text{(B.7)}$$

Since the subplanetary point lies in the orbital plane of the planet, it will be at a position ($\nu_{*\rm p}(t)$,0) on the planetary surface, where

$$\nu_{*\rm p}(t) = \arccos\left( \frac{\cos(E_{*\rm p}(t)) - e_{*\rm p}}{1 - e_{*\rm p}\cos(E_{*\rm p}(t))} \right) \qquad \text{(B.8)}$$

is the true anomaly. Moreover, with $s_{\rm x}$, $s_{\rm y}$, and $s_{\rm z}$ as the components of $\vec{s} = (s_{\rm x}, s_{\rm y}, s_{\rm z})$ we have

$$\begin{aligned} \Phi(t) &= 2\arctan\left( \frac{s_{\rm y}(t)}{\sqrt{s_{\rm x}^2(t) + s_{\rm y}^2(t)} + s_{\rm x}(t)} \right) \\ \vartheta(t) &= \frac{\pi}{2} - \arccos\left( \frac{s_{\rm y}(t)}{\sqrt{s_{\rm x}^2(t) + s_{\rm y}^2(t) + s_{\rm z}^2(t)}} \right) \end{aligned} \qquad \text{(B.9)}$$

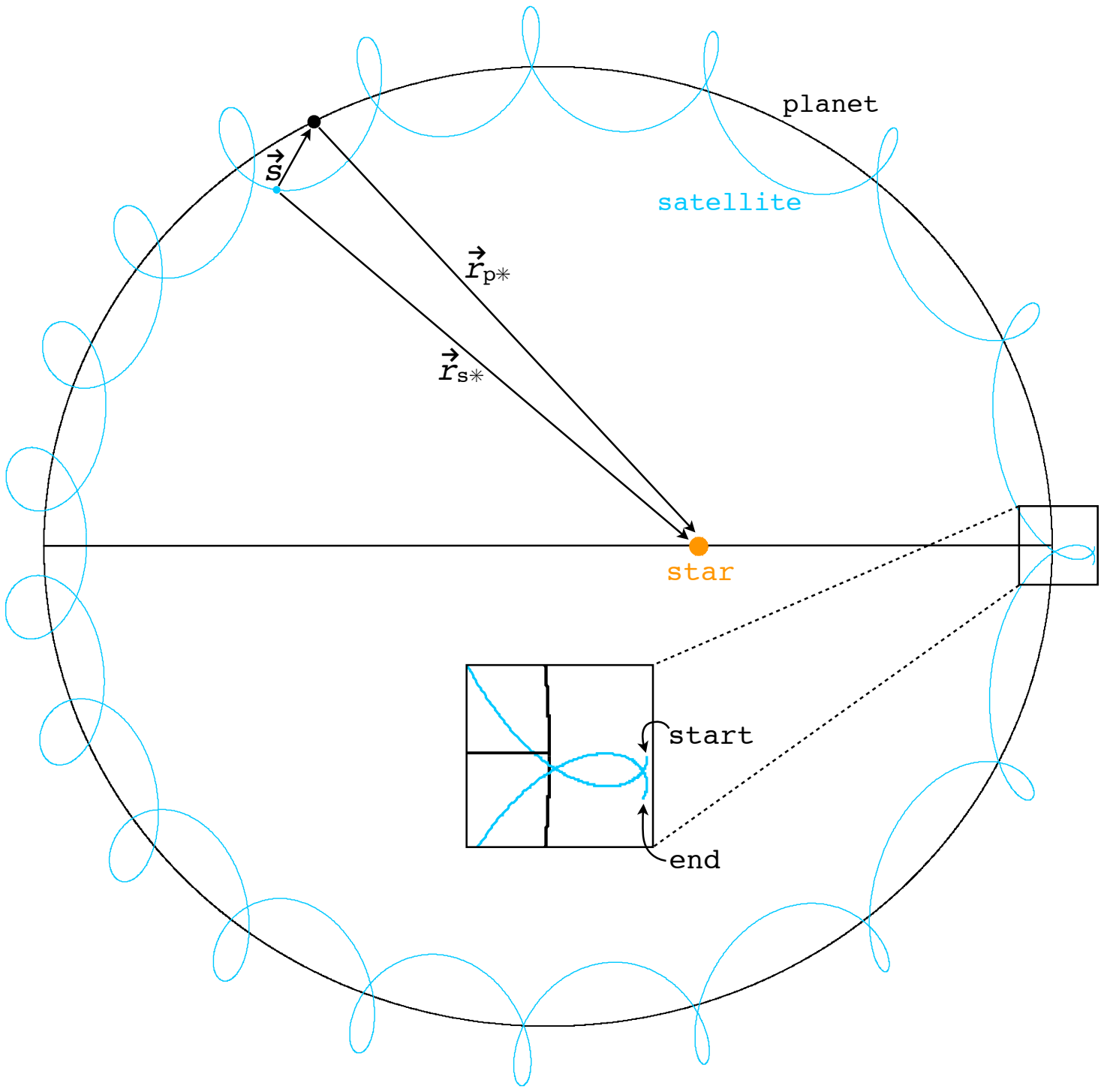

**FIG. C.1.** Orbits of an exomoon and an exoplanet around their common host star as computed with our `exomoon.py` software.

and thus determined $f_t(t)$.

We now consider the reflected stellar light from the planet. For the derivation of $f_r(t)$, we assumed that the planet is divided in two hemispheres, one of which is the bright side and one the dark side. Now the bright part coincides with the hemisphere from which the planet receives stellar reflected light, while there is no contribution to the reflectance from the dark side. Thus, the deduction of the geometrical part of $f_r(t)$, represented by $\xi(t)$, goes analogously to $f_t(t)$. We only have to multiply the stellar flux received by the planet $R_*^2 \sigma_{BS} T_{eff,*}^4 / r_{p*}^2$ with the amount $\pi R_p^2 \alpha_p$ that is reflected from the planet and weigh it with the squared distance decrease $a_{ps}^{-2}$ between the planet and its satellite. Consideration of projectional effects of longitude and latitude yields

$$f_r(t) = \frac{R_*^2 \sigma_{SB} T_{eff,*}^4}{r_{p*}^2} \frac{\pi R_p^2 \alpha_p}{a_{ps}^2} \times \cos\left(\frac{\phi\pi}{180°}\right) \cos\left(\frac{\theta\pi}{180°}\right) \xi(t) \ . \tag{B.10}$$

In Section 3.1.2, we have compared thermal radiation and stellar reflection from the planet as a function of the planet-moon distance $a_{ps}$ and the planet's albedo $\alpha_p$ (see Fig. 2). To compute the amplitudes of $f_t$ and $f_r$ over the moon's orbit around the planet, we assume that the moon is over the substellar point on the planet where the satellite receives both maximum reflected and thermal radiation from the planet. Then $\xi(t) = 1$ and equating Eq. (B.4) with (B.10) allows us to compute that value of $\alpha_p$ for which the two contributions will be similar. We neglect projectional effects of longitude and latitude and derive

$$\alpha_p = \frac{1}{\pi} \left(\frac{r_{p*}}{R_*}\right)^2 \left(\frac{T_{eff,p}^b}{T_{eff,*}}\right)^4_{|f_t = f_r} \ . \tag{B.11}$$

For our prototype system this yields $\alpha_p = 0.093$. For higher planetary albedo, stellar reflected light will dominate irradiation on the moon.

## Appendix C: Computer code of our model: `exomoon.py`

Finally, we make the computer code `exomoon.py`, which we set up to calculate the phase curves (Figs. 4 and 5) and surface maps (Figs. 7, 11, and 12), publicly available. It can be downloaded from www.physics.mcmaster.ca/~rheller or requested via email. The code is written in the programming language `python` and optimized for use with `ipython` (Pérez & Granger 2007). Care has been taken to make it easily human-readable, and a downloadable manual is available, so it can be modified by non-expert users. The output format are ascii tables, which can be accessed with `gnuplot` and other plotting software.

In brief, the program has four operation modes, which allow the user to compute (*i*.) phase curves of $f_*(\varphi_{ps})$, $f_r(\varphi_{ps})$, and $f_t(\varphi_{ps})$, (*ii*.) orbit-averaged flux maps of exomoon surfaces, (*iii*.) the orbit of the planet-moon duet around their common host star, and (*iv*.) the runaway greenhouse flux $F_{RG}$. In Fig. C.1, we show an example for such an orbit calculation. The moon's orbit is inclined by 45° against the plane spanned by the planet and the star, and the stellar orbit has an eccentricity $e_{*p} = 0.3$. Near periastron (at the right of the plot), where $\mathfrak{M}_{*p} = 0$, the orbital velocity of the planet-moon duet is greater than at apastron. This is why the moon's path is more curly at the left, where $\mathfrak{M}_{*p} = \pi$. Note that the starting point and the final point of the moon's orbit at the very right of the figure do not coincide! This effect induces TTVs of the planet.

**Acknowledgements**

René Heller received funding from the Deutsche Forschungsgemeinschaft (reference number schw536/33-1). Rory Barnes acknowledges support from the NASA Astrobiology Institute's Cooperative agreement No. NNH05ZDA001C that supports the Virtual Planetary Lab, as well as NSF grant AST-1108882. This work has made use of NASA's Astrophysics Data System Bibliographic Services and of Jean Schneider's exoplanet database (www.exoplanet.eu). Computations have been performed with `ipython 0.13` on `python 2.7.2` and figures have been prepared with `gnuplot 4.4` (www.gnuplot.info) as well as with `gimp 2.6` (www.gimp.org). We thank the two anonymous referees for their expert counsel.